# Water Production Activity of Nine Long-Period Comets from SOHO/SWAN Observations of Hydrogen Lyman-alpha: 2013-2016

Short Title: Water Production in Comets 2013-2016

M.R. Combi[1], T.T. Mäkinen[2], J.-L. Bertaux[3], E. Quémerais[3], S. Ferron[4], M. Avery[1] and C. Wright[1]

[1]Dept. of Climate and Space Sciences and Engineering

University of Michigan

2455 Hayward Street

Ann Arbor, MI 48109-2143

*Corresponding author: mcombi@umich.edu

[2]Finnish Meteorological Institute, Box 503

SF-00101 Helsinki, FINLAND

[3] LATMOS/IPSL

Université de Versailles Saint-Quentin

11, Boulevard d'Alembert, 78280, Guyancourt, FRANCE

[4]ACRI-st, Sophia-Antipolis, FRANCE

36 Pages

3 Tables

10 Figures

9 Supplementary Tables

Submitted to Icarus

**Abstract**

Nine recently discovered long-period comets were observed by the Solar Wind Anisotropies (SWAN) Lyman-alpha all-sky camera on board the Solar and Heliosphere Observatory (SOHO) satellite during the period of 2013 to 2016. These were C/2012 K1 (PanSTARRS), C/2013 US10 (Catalina), C/2013 V5 (Oukaimeden), C/2013 R1 (Lovejoy), C/2014 E2 (Jacques), C/2014 Q2 (Lovejoy), C/2015 G2 (MASTER), C/2014 Q1 (PanSTARRS) and C/2013 X1 (PanSTARRS). Of these 9 comets 6 were long-period comets and 3 were possibly dynamically new. Water production rates were calculated from each of the 885 images using our standard time-resolved model that accounts for the whole water photodissociation chain, exothermic velocities and collisional escape of H atoms. For most of these comets there were enough observations over a broad enough range of heliocentric distances to calculate power-law fits to the variation of production rate with heliocentric distances for pre- and post-perihelion portions of the orbits. Comet C/2014 Q1 (PanSTARRS), with a perihelion distance of only ~0.3 AU, showed the most unusual variation of water production rate with heliocentric distance and the resulting active area variation, indicating that when the comet was within 0.7 AU its activity was dominated by the continuous release of icy grains and chunks, greatly increasing the active sublimation area by more than a factor of 10 beyond what it had at larger heliocentric distances. A possible interpretation suggests that a large fraction of the comet's mass was lost during the apparition.

Keywords: Comets, coma; Ices; Ultraviolet observations; Comets: C/2012 K1 (PanSTARRS), C/2013 US10 (Catalina), C/2013 V5 (Oukaimeden), C/2013 R1 (Lovejoy), C/2014 E2 (Jacques), C/2014 Q2 (Lovejoy), C/2015 G2 (MASTER), C/2014 Q1 (PanSTARRS), C/2013 X1 (PanSTARRS)

Introduction

The Solar Wind ANisotropies (SWAN) instrument on board the SOlar and Heliospheric Observatory (SOHO) satellite has been operating in a halo orbit around the Earth-Sun L1 Lagrange point since shortly after its launch on 2 December 1995. During that time it has recorded the hydrogen Lyman-alpha comae of over 70 comets, as well as multiple apparitions each of several short-period comets that have returned more than once since 1995; water production rates have been determined from these observations. Data from 63 comet apparitions observed from 1996 to 2012 have recently been archived in and certified by the NASA Planetary Data System Small Bodies node (Combi, 2017). A number of comets have been observed in the SWAN full-sky survey data since 2012, and here we present the results of water production rates calculated from nine long-period comets observed from 2013 through 2016. For most of the comets we derive pre- and post-perihelion power-law fits to the production rate variation with heliocentric distance. We also discuss each comet in the context of already published results.

**Observations and Basic Model Analysis**

While the SWAN instrument on the SOHO spacecraft was designed primarily to observe the interplanetary hydrogen distribution in the whole sky and monitor how it is affected by the changing solar wind, it was planned from the beginning that it would be an excellent instrument for observing comets (Bertaux et al., 1998). From its place in a halo orbit around the Sun-Earth L1 Lagrange point, most of the sky except for regions fairly close to the Sun and that obscured by the spacecraft itself, which also mostly blocks the Earth, is seen by SWAN. Unlike ground-based observatories, which have difficulty observing comets close to the local sunrise or sunset horizon or those in the opposite hemisphere, SWAN can observe most comets that are bright enough, typically brighter than magnitude 10-12, most of the time, providing an excellent place to observe comets continuously over large portions of their orbits.

SWAN consists of a 5 x 5 array of 1-degree detectors that are scanned across the whole sky each day, creating a full sky map of hydrogen Lyman-alpha. Except for a special set of comet-specific observations of the Rosetta target comet 67P/Churyumov-Gerasimenko in 2009 (Bertaux et al., 2014), comets since 2007 have been observed only in the full sky images. This is also true for the comet results presented in this paper. Water production rates derived from a set of 65 apparitions of 50 comets (Combi, 2017) have recently been archived and certified in the Planetary Data System (PDS) Small Bodies Node (SBN) and can be found at URL https://pds-smallbodies.astro.umd.edu/holdings/pds4-soho:swan_derived-v1.0/SUPPORT/dataset.html. The results contained in the paper include comets observed after this first set, though these data will be archived in the PDS at some time.

The SWAN instrument and its use for observing comets has been discussed by Bertaux et al. (1998) and Combi et al. (2005), and the updated calibration of the instrument for comets was discussed by Combi et al. (2011a). Since atomic hydrogen in comets is produced mostly in the photodissociation chain of $H_2O$ and OH it is possible to determine the water production rate in comets from observations of the hydrogen Lyman-alpha coma. The so-called time-resolved model used to calculate water production rates was presented and discussed in detail by Mäkinen and Combi (2005) and the determination of the various model parameters and parameterizations have been discussed in the Comets II review chapter by Combi et al. (2004). The fluorescence rate, or g-factor, of hydrogen atoms illuminated by the Sun is determined from the daily average solar irradiance compiled at LASP of the University of Colorado at the URL

http://lasp.colorado.edu/lisird/lya/. Because comets do not see the same face of the Sun as do the near-Earth satellites that contribute to the LASP solar measurements, we take the value from the nearest date corresponding to the heliographic longitude differences between the Earth and comet.

Nine long-period or dynamically new Oort cloud comets were observed during the period of 2013-2016 by SWAN. Orbital characteristics, time periods, and numbers of images are listed in Table 1. Table 1 contains the value of the original semi-major axis of the orbit ($a_0$) for each comet obtained from the Minor Planet Center (http://www.minorplanetcenter.net/db_search/). In the last column we have adopted the dynamical class scheme from the 85 comets photometric survey of A'Hearn et al. (1995) where dynamically new comets (DN) have semi-major axes larger than 20000 AU, Young Long-Period comets (YL) have semi-major axes between 500 and 2000 AU and Old Long-Period comets (OL) have semi-major axes less than 500 AU. It is estimated that ~90% of DN could be making their first trip into the inner solar system whereas most of the YL and OL comets have probably been to the planet region of the solar system before. The JPL Horizons web site was also used to estimate the original semi-major axis values. Most of the values were the same except for the DN comets of very large $a_0$, but in all cases the classification assignments were the same.

Complete tables of water production rates are included with this paper as Supplementary information because of their length. Tables S1 through S9 provide observational parameters, water production rates and random stochastic 1-sigma uncertainties associated with noise in the data and the model fitting procedure that accounts for the subtraction of the H Lyα emission from the interplanetary background, which is typically on the order of 1 kilorayleigh. For most of the comets water production rates for both pre- and post-perihelion parts of the orbit have been expressed as power laws as a function of heliocentric distance in the form $Q = Q(1AU) r^a$, where Q is the water production, Q(1AU) is the water production rate at a heliocentric distance of 1 AU, and r is the heliocentric distance of the comet in AU. Those results for all 9 comets are given in Table 2. In the following discussion section we will show the results for each comet and discuss the various implications and compare with other observations when available. When comparing water production rates determined from SWAN measurements of H Lyα emission we have estimated that there is a possible systematic uncertain of ~30% owing to a combination of instrument calibration, the model parameterization and the various model parameters.

**Results of Model Analyses for Each Comet**

C/2012 K1 (PanSTARRS). The H Lyα coma of comet C/2012 K1 (PanSTARRS) was observed by SWAN from 22 April to 6 December 2014 covering a range of heliocentric distances from 2.375 AU before perihelion to 1.912 AU after. It reached a perihelion distance of 1.0545 AU on 27.7 August 2014. Its original semi-major axis was 55000 AU. This, combined with its activity shallow production slope before perihelion and it normal behavior after perihelion, would indicate that it is likely a new Oort cloud comet on its first passage through the inner solar system. The results are given in Table S1. Figures 1 a & b show the water production rate as a function of time and heliocentric distance, respectively, over the apparition. The power-law variation of water production rate before and after perihelion shown as the solid lines in Figure 1b are given by $2.0 \times 10^{29} r^{-0.8}$ molecules s$^{-1}$ and $1.9 \times 10^{29} r^{-2.4}$ molecules s$^{-1}$, respectively.

In some ways the variation of production rate is reminiscent of comet C/2009 P1 (Garradd), which was also dynamically new, with a flat but very irregular and unusual variation on the inbound pre-perihelion leg inside 2.1 AU and a more normal and rather steady $r^{-2.4}$ drop after perihelion. Like Garradd (Combi et al., 2013; Bodewits et al., 2014), C/2012 K1 (PanSTARRS) is suspected of having an important icy grain halo before perihelion that likely dominates the water production rate. McKay et al. (2016) suggest that there is evidence in the radial distribution of gas emission from 2012 K1 from a range of observations with different aperture sizes, showing that there is an important extended source of gas. Their results, derived from relatively large-aperture Swift/UVOT observations of OH, are consistent with both the much larger aperture SWAN results shown here and the ground-based OH observations by Knight and Schleicher (2014), but they see a trend when comparing production rates from much smaller aperture observations.

C/2013 US10 (Catalina). SWAN observations of the Ly$\alpha$ coma of comet C/2013 US10 (Catalina) from 30 June 2015 to 18 February 2016 yielded water production rates of the comet corresponding to heliocentric distances of 2.12 AU before perihelion to 1.91 AU after. The comet reached its perihelion distance of 0.823 AU on 15 November 2015. Its original semi-major axis, like that of C/2012 K1 (PanSTARRS), was quite large at 14000 AU, which is just below the cut-off of the dynamically new class, and indicates that it may not have been on its first pass through the inner solar system. Its activity variation, as indicated by the power-law slopes, are only somewhat flatter than the canonical -2, as shown in Figure 2a and as a function of heliocentric distance in Figure 2b. The results are given in Table S2. The variation with heliocentric distance can be described with a power-law slope of -1.6 before perihelion and -1.9 after perihelion and an otherwise steady increase and decrease with no unusual irregular variations or outbursts. Water production rates are generally higher before perihelion than after.

C/2013 V5 (Oukaimeden). Comet C/2013 V5 (Oukaimeden) was discovered as part of the Morocco Oukaimeden Sky Survey (MOSS) carried out by the observatory in Oukaimeden, Morocco on 12 November 2013 (Guido et al., 2013). The comet reached a perihelion distance of 0.625 AU on 28 September 2014. SWAN obtained images of its H Ly$\alpha$ coma from 19 August 2014 at a heliocentric distance of 1.01 AU to 7 October 2014, only a week after perihelion as its production rate and brightness were dropping much more rapidly than the very flat rise before perihelion. It also appears to be a dynamically new comet from the Oort cloud. The water production rate is shown plotted as a function of time from perihelion in Figure 3. The results are given in Table S3. While there were not enough data after perihelion to describe the variation with a power-law fit, we were able to fit a power law to the pre-perihelion data $3.4 \times 10^{28} r^{-1.0}$ molecules s$^{-1}$. Comet C/2013 V5 (Oukaimeden), like C/2012 K1 (Pan STARRS), had an original semi-major axis of 55000 AU, putting it easily into the dynamically new group. The very flat slope is consistent with it being a dynamically new comet on its first passage into the inner solar system.

C/2013 R1 (Lovejoy). SWAN obtained useful images of the H Ly$\alpha$ coma of comet C/2013 R1 (Lovejoy) from 13 October 2013 to 14 March 2014. The comet reached its perihelion of 0.812 AU on 22 December 2013. It has an original semi-major axis of 365 AU, so it does not appear to be a dynamically new comet on its first pass from the Oort cloud. Figures 4a and 4b give the water production rates as a function of time from perihelion and heliocentric distance, r, that are also listed in Table S4. Power laws of the forms $7.6 \times 10^{28} r^{-2.2}$ molecules s$^{-1}$ and $9.1 \times 10^{28} r^{-1.6}$ molecules s$^{-1}$ can describe the variation with time.

Water production rates determined by Biver et al. (2014) from OH observations from the Institut de Radioastronomie Millimétrique (IRAM) at pre-perihelion heliocentric distances of 1.13, 0.92, and 0.83 are in excellent agreement with the SWAN derived values at the same times. Biver et al. also reported IRAM detections of complex organics ethylene glycol and formamide in comet 2013 R1 (Lovejoy) at the 0.02% level compared with water.

On the other hand, water production rates determined from water IR emissions with NIRSPEC at the Keck Observatory by Paganini et al. (2014) over the pre-perihelion heliocentric distance range from 1.35 to 1.16 AU are systematically a factor of 2 or more below the SWAN values and the one overlapping value from Biver et al. (2014). Similarly the ground-based near-ultraviolet OH production rates given by Opitom et al. (2015), after accounting for the ~0.82 OH yield from $H_2O$ photodissociation (Combi et al., 2004), like the NIRSPEC/Keck observations given by Paganini et al. (2014), are systematically smaller by factors of 2 or more from the SWAN and IRAM rates.

Figure 4b also shows various water production rate measures for comet C/2013 R1 (Lovejoy) determined from different observations. Opitom et al. (2015) have adopted the OH Haser scale lengths and methods from A'Hearn et al. (1995) to calculate OH production rates. Schleicher and Osip (2002) has shown that in order to compare water production determined from OH observations with those of other methods that the production rates need to be corrected for the more appropriate vectorial model (Festou, 1981), which is more typically used by space-based ultraviolet observations of OH (e.g., Swift by Bodewits et al., 2014). A number of comparisons of water production rates determined from SWAN observations of atomic hydrogen have shown that the vectorial model approach normally corrects for this factor of ~2 (Combi et al., 2000, 2005). Therefore the factor of ~2 offset between the water production rates derived from Opitom et al. and SWAN, determined here for the pre-perihelion period, is not surprising, but the factor of nearly 4 offset for the post-perihelion period is quite puzzling. It does appear that the variation with heliocentric distance between the two datasets is consistent during the periods of overlap.

In the case of 2009 P1 (Garradd) there was a correlation between observational aperture size and calculated water production rate that was consistent with the interpretation that much of the water production resulted from a spatially extended sublimation source of icy particles (Combi et al., 2013; Bodewits et al., 2014). However, 2009 P1 (Garradd) was a dynamically new comet while 2013 R1 (Lovejoy) was not. While the aperture size for the ground-based IR measurements (Paganini et al., 2017) are the smallest, the effect of aperture size from the ground-based OH observations are intermediate and not much different from the IRAM measurements (Biver et al., 2016), even though the IRAM water production rates are consistent with the SWAN determined results having by far the largest observational aperture. The explanation may have to await the publication of more results.

C/2014 E2 (Jacques). Like C/2013 R1 (PanSTARRS), C/2014 E2 (Jacques) appears to be a very long period comet, but is probably not on its first pass from the Oort cloud, given its original semi-major axis of only 365 AU. SWAN obtained useful detections of this comet over six months of its apparition from 3 April 2014 to 19 October 2014, and heliocentric distances from 1.8 AU before perihelion to 2.1 AU after. The variation with heliocentric distance was rather "normal" having an exponential slope of -2.4 before perihelion and -1.7 after, and a small asymmetry in activity with somewhat higher water production rates before than after perihelion, on the average. The largest measured production rate occurred about four weeks before perihelion, however, there was a long period on either side of perihelion when the comet was not observable. Figures 5a and 5b show the variation of water production rate with time and with heliocentric distance, respectively. The numerical values are listed in Table S5. Figure 5b also shows the best fit power law with heliocentric distance, r in AU, expressed as $1.5 \times 10^{29}$ $r^{-2.4}$ molecules s$^{-1}$ and $1.1 \times 10^{29}$ $r^{-1.7}$ molecules s$^{-1}$, respectively.

C/2014 Q2 (Lovejoy). Like 2013 R1 and 2014 E2 C/2014 E2 (Jacques), comet C/2014 Q2 (Lovejoy) is a very long period but not a new comet. Six months' worth of observations were obtained from SWAN full-sky images covering the period from 28 November 2014 to 28 June 2015 and including a perihelion distance of 1.29 AU on 30 January 2015. Heliocentric distances covered were from 1.58 AU before perihelion to 2.17 AU after.

The variation of water production over time is shown in Figure 6a. Despite the fact that most returning bright comets from the Oort cloud on long-period orbits observed by SWAN over the years tend to have rather typical variations with heliocentric distance and power-law slopes in the range from -1.5 to -2.5, the variation of water production of comet C/2014 Q2 (Lovejoy) is quite unusual. The peak production rate of $8 \times 10^{29}$ molecules s$^{-1}$ occurred about 20 days after perihelion and production rates were generally much larger after perihelion than before. The power-law fit to the pre-perihelion data shown in Figure 6b was also rather unusual, having rather two different components. From a heliocentric distance of 1.4 AU to perihelion the slope was a very steep -6.6, and in calculating the slope these data dominated the fit. However, for distances from 1.6 to 1.4 AU before perihelion the variation was more irregular and less steep with a slope closer to -3, which is more similar to the generally consistent post-perihelion slope of -3.4. The numerical results are listed in Table S6. The original semi-major axis of 500 AU, its irregular activity behavior, both combined with its strong seasonal effect, imply that it is consistent with having a nucleus that is quite evolved after having had many passages into the inner solar system. Other than having a very long period and a large absolute production rate at 1 AU, its behavior is more like many Jupiter family comets.

In an important paper by Biver et al. (2015) showing the detection of 21 molecules, including ethyl alcohol and sugar, they find a water production rate of $(5.0\pm0.2) \times 10^{29}$ molecules s$^{-1}$ from observations of OH made with the Nançay radio telescope during the few days around perihelion and $(7.5\pm0.3) \times 10^{29}$ molecules s$^{-1}$ of $H_2O$ made with the Odin satellite by Biver et al. (2016). From the combination they estimate the water production rate just before perihelion of $6.0 \times 10^{29}$ molecules s$^{-1}$. These are consistent with the SWAN measurements during the same period.

Paganini et al. (2017) have reported observations of $H_2O$ and HDO obtained with NIRSPEC at the Keck Observatory on 4 February 2015, just 1 day after the end of the period covered by Faggi et al. (2016), and report a water production rate of $(5.0 \pm 0.13) \times 10^{29}$ molecules s$^{-1}$, also in agreement with the values of SWAN to well within the expected uncertainty ranges.

Faggi et al. (2016) observed comet C/2014 Q2 (Lovejoy) on the 3 days following perihelion with the Giano spectrograph on the Telescopio Nazionale Galileo (TNG) telescope at LaPalma, Canary Islands, and found a water production rate of $(3.11\pm0.14) \times 10^{29}$ molecules s$^{-1}$. This appears systematically lower than the other measurement but is technically within the SWAN results when accounting for the possible ±30% systematic uncertainty from various model parameters and calibration. The results of Faggi et al. (2016) and Paganini et al. (2017) are indicated on Figure 6a.

Finally, unpublished results from the TRAPPIST observatory (Jehin, private communication) give a ground-based OH production rate on 7 January 2015 of $(2.71\pm0.30)$ x $10^{29}$ molecules s$^{-1}$, which implies a water production rate of 3.30 x $10^{29}$ when dividing by the OH branching ratio of 0.82. This is compared with the SWAN value on the same day of 2.87 x $10^{29}$ molecules s$^{-1}$. These are within their respective uncertainty estimates.

C/2015 G2 (MASTER). SWAN obtained useful observations of comet C/2015 G2 (MASTER) from 7 April 2015 to 26 June 2015. The comet reached a perihelion distance of 0.78 AU on 23 May 2015. Given its orbit, it is likely to be a dynamically new comet from the Oort cloud. The water production rates plotted as a function of time in Figure 7 indicate a very irregular variation over its orbit. See Table S7 for the numerical results and observational aspects. A power law in heliocentric distance, r, can be fitted to the pre-perihelion variation, yielding an unremarkable production rate variation of 4.1 x $10^{28}$ r$^{-1.8}$ molecules s$^{-1}$. A conventional power law following perihelion does not seem to have any meaning as the production rate is simply irregular and very flat over that period.

There are not as yet any published water production rates for C/2015 G2 (MASTER). Jehin (private communication) has reported an OH production rate on 21 April 2015, which when converted as above to a water production rate yields a value of 3.26 x $10^{28}$ molecules s$^{-1}$ that compares favorably to the SWAN value on the same day of 3.86 x $10^{28}$ molecules s$^{-1}$. The original semi-major axis of this comet is 30400 AU, which puts it into the dynamically new category. While its inbound slope is not much flatter than -2, its very irregular variation both pre- and post-perihelion could be consistent with a first-time passage into the inner solar system.

C/2014 Q1 (PanSTARRS). The apparition of comet C/2014 Q1 (PanSTARRS) was quite short, covering only 90 days from 20 May 2015 through 10 August 2015. During that time the water production rate rose very rapidly to a value of almost $2 \times 10^{30}$ molecules s$^{-1}$ a few days before its relatively short perihelion distance of 0.32 AU from the Sun on 6 July 2015. It also declined similarly and very rapidly after perihelion. The water production rate, as determined from the SWAN observations of the H Ly$\alpha$ coma, is plotted as a function of time from perihelion in Figure 8. The numerical values are given in Table S8. It appears to be a very long period comet but not dynamically new on its first trip to the inner solar system from the Oort cloud, having an original semi-major axis of 825 AU. Its production rate can be well described by two power laws of the form $5.3 \times 10^{26}$ $r^{-7.8}$ molecules s$^{-1}$ pre-perihelion and $9.1 \times 10^{26}$ $r^{-8.9}$ molecules s$^{-1}$ post-perihelion.

The very steep rise near the very small heliocentric distances is reminiscent of comet C/2012 S1 (ISON), which began an extremely rapid increase in activity about 15 days before perihelion at a heliocentric distance of 0.70 AU, at which time its active area increased markedly compared with the more normal constant active area before that time (Combi et al., 2014). The interpretation for comet C/2012 S1 (ISON) is that its small nucleus began to disrupt, shedding icy grains and chunks at an increasing rate until its demise at its extremely small perihelion distance of less than 2 solar radii. If one only considers the variation of water production for heliocentric distances larger than ~0.7 AU, that is more than about 25 days on either side of perihelion, like ISON, the variation of C/2014 Q1 (PanSTARRS) is somewhat irregular, but rather flat.

Not all comets with very small perihelion distances exhibit this kind of behavior. A study of four comets with small perihelia, namely C/2002 V1 (NEAT), C/2002 X5 (Kudo–Fujikawa), 2006 P1 (McNaught) and 96P/Machholz 1, of 0.099, 0.190, 0.170 and 0.13 AU, respectively, all had reasonably normal variations with heliocentric distance with power-law slopes in the range of -1.5 to -3.5, pre- and post-perihelion (Combi et al., 2011b). So simply having a comet move very close to the Sun does not initiate this intense level of activity that greatly exceeds that which might be expected from some combination of a moderate seasonal effect in combination with water-dominated sublimation. This would imply that there was something different in the structure of comets C/2012 S1 (ISON) and C/2014 Q1 (PanSTARRS) than many other comets that pass well within 0.7 AU from the Sun but continue to sublimate steadily.

Circumstantial evidence might suggest that C/2014 Q1 (PanSTARRS) is probably similar to C/2012 S1 (ISON), and that it similarly began to rapidly shed chunks of nucleus material once it was closer to the Sun than about 0.7 AU, at which time the water production began to increase far more rapidly than something close to a constant cross-sectional area exposed to sunlight can. The fact that the post-perihelion activity level returned to a similar but slightly lower level once the comet returned to 0.7 to 1.0 AU suggests, however, that, unlike C/2012 S1 (ISON), a reasonable fraction of the nucleus remained intact after the apparition. This was also aided by the fact that the perihelion distance was not nearly as close as C/2012 S1 (ISON).

Figure 9 shows a plot of the active area of comet C/2014 Q1 (PanSTARRS) calculated following the method of Cowan and A'Hearn (1979), as done for comet C/2012 S1 (ISON) by Combi et al. (2014). It shows two relatively steady regions before 20 days before perihelion and after 20 days after perihelion, with up to 100 km$^2$ during the period around perihelion. The average active area of the steady region well before perihelion, shown as the thick solid line on the left part of the plot, has a value of 9.6 km$^2$, which is about 1.5 times larger than the value of 6 km$^2$ found for comet C/2012 S1 (ISON) well before perihelion. On the right side of the plot the average active area, once it returns to another relatively steady level, is 4.9 km$^2$. From this we can conclude that the comet lost about half of its material during the very active phase near perihelion when it was shedding material with a total surface area of up to six times the original surface area of the nucleus.

The total water mass released during the whole observed apparition of ±40 days from perihelion is 3.1 x 10$^{11}$ kg, with about 90% of this released from the enhanced active area peak of ±25 days from perihelion and about 10% from what would be the nucleus itself as determined by the flatter steady release before and after 25 days from perihelion. For comparison, this is a hundred times more than comet 67P/Churyumov-Gerasimenko, which lost about 3 x 10$^9$ kg during each orbit in 1996, 2002 and 2007 (Bertaux, 2015). This is also far larger than the mass comet C/2012 S1 (ISON) lost with the complete disruption of its nucleus (Combi et al., 2014). If the material in the nucleus of C/2014 Q1 (PanSTARRS) is assumed to be homogeneous, the reduction of active area would imply a decrease of the nucleus radius to 71% of its original size and a loss of 64% of its original total mass. The actual size of the nucleus will depend on the ratio of the water mass to the total mass. If we assume the same bulk nucleus density as 67P/Churyumov-Gerasimenko from Rosetta measurements (Pätzold et al., 2016) of 533 kg/m$^3$, we can estimate the radius of the nucleus before and after perihelion for a few values of the water-to-total mass ratio.

These are given in Table 3 for values of the water to total mass ratio of 1, 1/2, 1/3 and 1/4. This ratio roughly corresponds to the inverse of the dust-to-gas mass ratio. For this fairly wide range of values, we find nucleus radii before the peak of the apparition in the range of 725 to 1151 meters. This indicates the nucleus of C/2014 Q1 (PanSTARRS) was quite a bit larger than that of C2012 S1 (ISON), which completely disintegrated on its extremely close pass by the Sun, however it did lose most of its mass on this passage.

C/2013 X1 (PanSTARRS). Comet C/2013 X1 (PanSTARRS) is a young long-period (YL) comet, having an original semi-major axis of 4450 AU, and is not likely on its first trip into the inner solar system from the Oort cloud. Its H Lyα coma was detected in 121 full-sky SWAN images from 1 January 2016 to 21 June 2016. It reached a perihelion distance of 1.31 AU on 20 April 2016. It reached a peak water production rate of 4.4 x $10^{29}$ molecules $s^{-1}$ 15 days before perihelion. Water production rates and observational geometry parameters are given in Table S9.

Water production rates are plotted in Figure 10a as a function of time from perihelion in days and with pre- and post-perihelion data plotted as a function of heliocentric distance in Figure 10b. Table S9 gives the numerical values, uncertainties and observational parameters. This comet has a general variation over its orbit that can be expressed as a power law in heliocentric distance, r in AU, as 8.2 x $10^{29}$ $r^{-3.0}$ molecules $s^{-1}$ before perihelion and 2.7 x $10^{29}$ $r^{-2.2}$ molecules $s^{-1}$ after. Throughout March 2014 just before perihelion the comet could not be observed as it was in the solar avoidance area of the sky. Just after it was detectable again, the production rate started what appears to be an outburst that peaked two days later. The outburst lasted about 12 days and returned to the pre-perihelion value of just over $10^{29}$ molecules $s^{-1}$ a few days before perihelion. SWAN has a fairly small solar avoidance angle so it is doubtful that there are any (or many) observations of the comet during the period just before the outburst. After perihelion the comet's production rate generally drops fairly steadily as heliocentric distance increases.

**Summary**

We presented the results of a model analysis of over 800 images of the hydrogen Lyman-alpha comae of 6 long-period and 3 dynamically new comets observed with the SWAN all-sky camera on the SOHO spacecraft during the period of 2012 to 2016. Water production rates were calculated from each of the images with our standard method (Mäkinen and Combi, 2005). Regarding water production rates and activity over the orbits of the comets observed, two of the three comets, C/2012 K1 (PanSTARRS) and C/2013 V5 (Oukaimeden), fell into the dynamically new class and displayed the rather flat variations on their inbound orbits that one might expect for comets on their first trips into the inner solar system. The overall slope of -1.8 of the inbound variation of the third, C/2015 G2 (MASTER), was not particularly flat, but the variation was highly irregular both pre- and post-perihelion, and the post-perihelion was so irregular that fitting a power law was not meaningful. The remaining six long-period comets showed a wide range of variations with two having very steep variations, -6.6 and -3.4 slopes for C/2014 Q2 (Lovejoy) and -7.8 and -8.9 for C/2014 Q1 (PanSTARRS) pre- and post-perihelion, respectively.

Where available we compared the SWAN water production rates with some obtained either by water directly in the IR or inferred from OH ground-based observations. For some comets and some datasets agreement is quite good, while for others there are sometimes apparent systematic differences that have been seen in previous comparisons of other comets.

The most remarkable was comet C/2014 Q1 (PanSTARRS), which at first look presented very steep average power-law slopes of -7.8 and -8.9 before and after perihelion, respectively. A closer examination of the variation of water production rate over time showed that the rapid rise was confined to a period of within ±25 days of perihelion while the comet was roughly within 0.7 AU from the Sun. Outside that period the activity was somewhat irregular but generally steady. Calculations of the active area using the standard method of Cowan and A'Hearn (1979) showed that the equivalent active area was nearly a factor of ten larger during the perihelion time ($r_H \sim 0.31$ AU) than both well before and after perihelion, and that the average active area well after perihelion was only half of the value well before perihelion. This behavior was generally similar to that of comet C/2012 S1 (ISON) covering the same heliocentric distances. However, both comets showed activity consistent with the shedding of large amounts of icy grains and chunks from their nuclei when they were within 0.7 AU from the Sun greatly increasing their active sublimation areas. While comet C/2012 S1 (ISON) did not survive intact its extremely small (< 2 solar radius) perihelion passage, comet C/2014 Q1 (PanSTARRS) did survive its more moderate 0.31 AU perihelion but had an active area about a factor of two smaller after perihelion than before, indicating that perhaps 64% of its original mass was lost, and its radius was reduced to 71% of its incoming value. Lastly, there are, of course, many comets that reach perihelion distances much smaller than 0.3 AU but exhibit normal activity (Combi et al., 2011a) and no evidence of greatly increased active surface area resulting from a violent release of icy grains and chunks from the nucleus.

**Acknowledgements**

We thank the two referees, Dr. Michael DiSanti and Dr. David Schleicher, for their careful reading and helpful comments, which have greatly improved this paper. SOHO is a cooperative international mission of ESA and NASA. M. Combi acknowledges support from grant NNX15AJ81G from the Solar System Observations Planetary Astronomy Program and NNX13AQ66G from the Planetary Mission Data Analysis Program. T.T. Mäkinen was supported by the Finnish Meteorological Institute (FMI). J.-L. Bertaux and E. Quémerais acknowledge support from CNRS and CNES. We obtained cometary ephemerides and orbital elements from the JPL Horizons web site (http://ssd.jpl.nasa.gov/horizons.cgi). We obtained original semi-major axis values for each of the observed comets from the IAU Minor Planet Center (http://www.minorplanetcenter.net/db_search/). The composite solar Lyman-alpha data were obtained from the LASP web site at the University of Colorado (http://lasp.colorado.edu/lisird/lya/). We also gratefully acknowledge the personnel that have been keeping SOHO and SWAN operational for over 20 years, in particular Dr. Walter Schmidt at FMI.

Table 1. Comets Observed with SWAN, Data Description and Orbital Parameters

| Comet | q(AU) | $a_0$ (AU) [a] | $r_H$ (AU) [b] | No. of images | Class [c] |
|---|---|---|---|---|---|
| C/2012 K1 (PanSTARRS) | 1.0545 27.7 Aug 2014 | 55340 | 2.197 - 1.055 1.055 - 1.912 | 138 | DN |
| C/2013 US10 (Catallina) | 0.8230 15.7 Nov 2015 | 14430 | 2.375 - 0.950 0.869 - 1.823 | 119 | YL |
| C/2013 V5 (Oukaimeden) | 0.6255 28.2 Sep 2014 | 55160 | 1.010 – 0.626 0.633 – 0.666 | 41 | DN |
| C/2013 R1 (Lovejoy) | 0.8115 22.7 Dec 2013 | 365 | 1.468 – 0.812 0.812 – 1.701 | 141 | OL |
| C/2014 E2 (Jacques) | 0.6639 2.5 Jul 2013 | 790 | 1.763 – 0.837 0.699 – 2.066 | 97 | YL |
| C/2014 Q2 (Lovejoy) | 1.290 30.1 Jan 2015 | 500 | 1.583 - 1.291 1.291 - 2.172 | 138 | OL/YL |
| C/2015 G2 (MASTER) | 0.7798 23.8 May 2015 | 30400 | 1.159 - 0.781 0.780 - 1.014 | 59 | DN |
| C/2014 Q1 (PanSTARRS) | 0.3146 6.5 Jul 2015 | 826 | 1.189 - 0.335 0.495 - 0.958 | 31 | YL |
| C/2013 X1 (PanSTARRS) | 1.3143 20.8 Apr 2016 | 4500 | 2.235 - 1.314 1.314 - 1.972 | 121 | YL |

a. The original semi-major axis before the perihelion passage.
b. For each comet the heliocentric distance range is given for pre- and post-perihelion periods.
c. Dynamical classes are DN (Dynamically New), OL (Old Long Period), YL (Young Long Period) as discussed in the text.

Table 2. Water Production Rate Heliocentric Distance Dependencies

| Comet | Q Pre-Perihelion[a] | Q Post-Perihelion[a] |
|---|---|---|
| C/2012 K1 (PanSTARRS) | $2.0 \times 10^{29}\ r^{-0.8}$ | $1.9 \times 10^{29}\ r^{-2.4}$ [1] |
| C/2013 US10 (Catalina) | $2.8 \times 10^{29}\ r^{-1.6}$ | $1.6 \times 10^{29}\ r^{-1.9}$ |
| C/2013 V5 (Oukaimeden) | $3.4 \times 10^{28}\ r^{-1.0}$ | - |
| C/2013 R1 (Lovejoy) | $7.6 \times 10^{28}\ r^{-2.2}$ | $9.1 \times 10^{28}\ r^{-1.6}$ |
| C/2014 E2 (Jacques) | $1.5 \times 10^{29}\ r^{-2.4}$ | $1.1 \times 10^{29}\ r^{-1.7}$ |
| C/2014 Q2 (Lovejoy) | $2.3 \times 10^{30}\ r^{-6.6}$ | $1.9 \times 10^{30}\ r^{-3.4}$ |
| C/2015 G2 (MASTER) | $4.1 \times 10^{28}\ r^{-1.8}$ | - |
| C/2014 Q1 (PanSTARRS) | $5.3 \times 10^{26}\ r^{-7.8}$ | $9.1 \times 10^{26}\ r^{-8.9}$ |
| C/2013 X1 (PanSTARRS) | $8.2 \times 10^{29}\ r^{-3.0}$ | $2.7 \times 10^{29}\ r^{-2.2}$ |

a. Production rate at 1 AU in molecules s$^{-1}$.

Table 3. Estimates of a Spherical Nucleus Radius Loss for C/2014 Q1 (PanSTARRS)

| Water mass fraction of the nucleus | f=1 | f=1/2 | f=1/3 | f=1/4 |
|---|---|---|---|---|
| Total mass loss ±40 days (kg) | $3.1 \times 10^{11}$ | $6.2 \times 10^{11}$ | $9.3 \times 10^{11}$ | $12.4 \times 10^{11}$ |
| Radius before -25 days (m) | 725 | 913 | 1045 | 1151 |
| Radius after +25 days (m) | 515 | 648 | 741 | 816 |

f = the water mass fraction of the nucleus

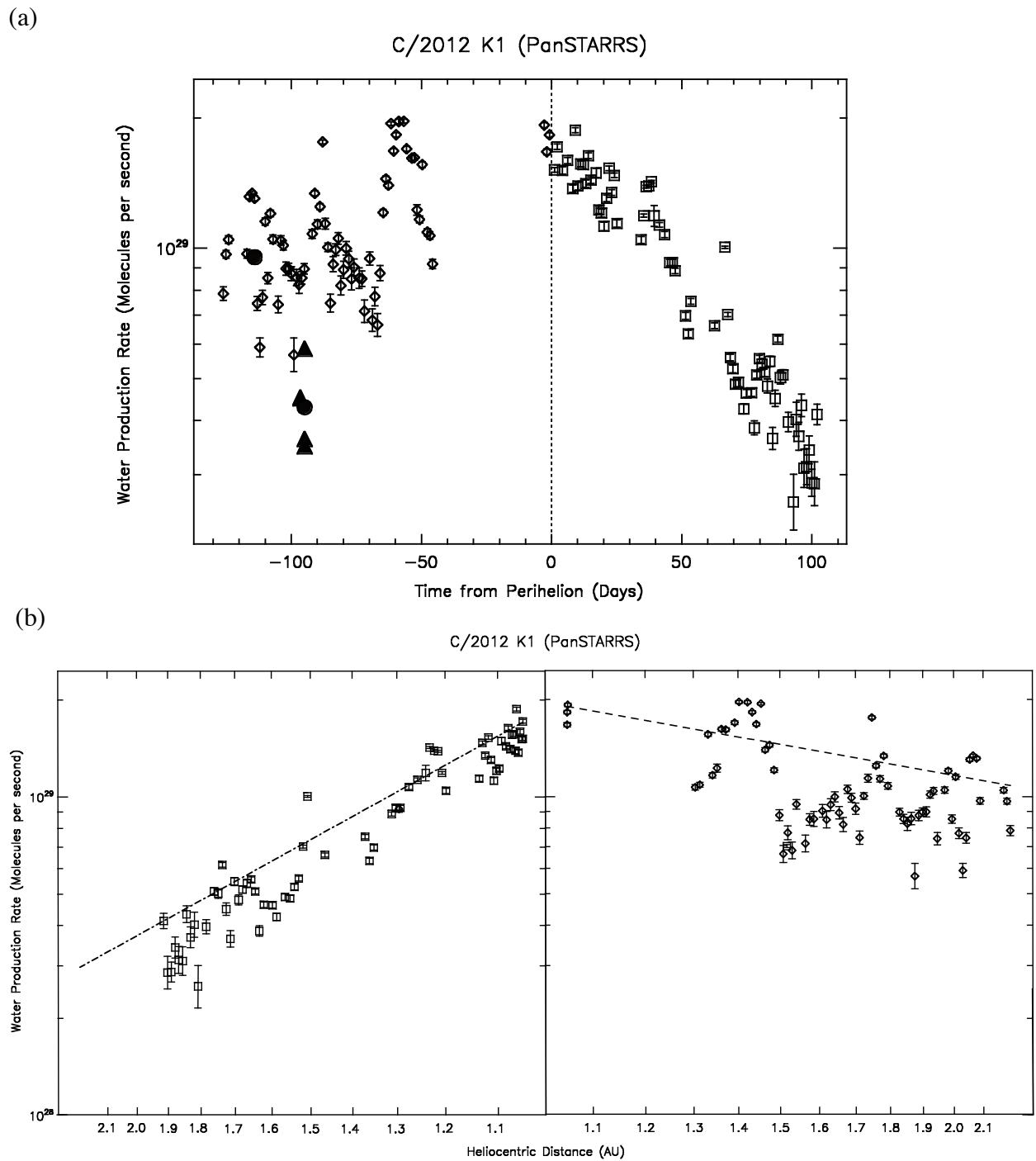

**Fig. 1.** Water production rate in comet C/2012 K1 (PanSTARRS) (a) shows the variation over time in days from perihelion, and (b) shows the variation with heliocentric distance with the diamonds giving the pre-perihelion values and the squares giving the post-perihelion values. The error bars correspond to the 1-sigma stochastic errors from noise in the data and the model fitting. The lines give the best-fit power-law variation with form given in Table 2. Shown in (a) are the measurements of McKay et al. (2016) as filled circles and of Roth et al. (2017) as filled triangles.

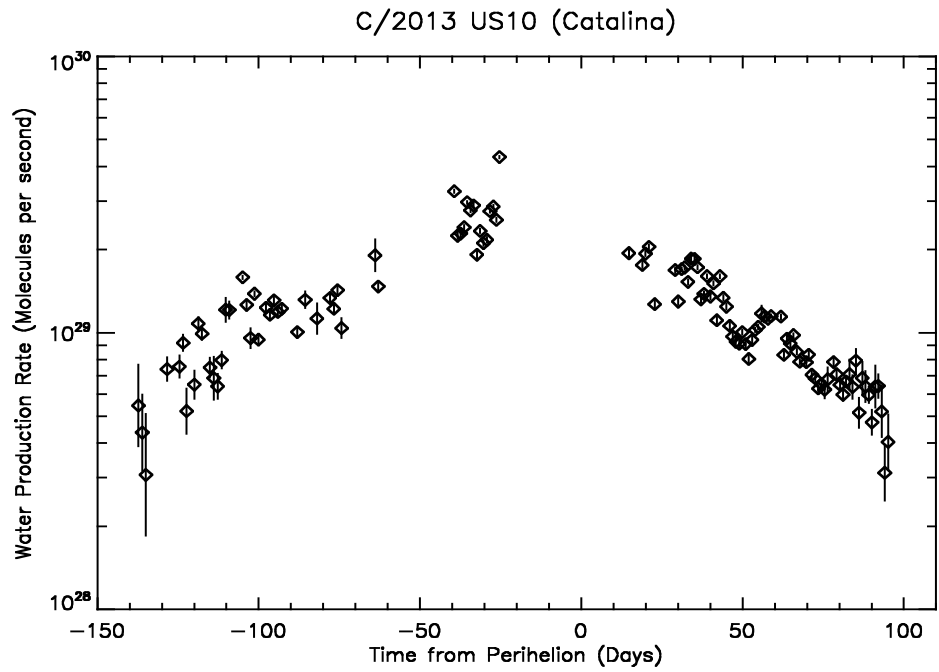
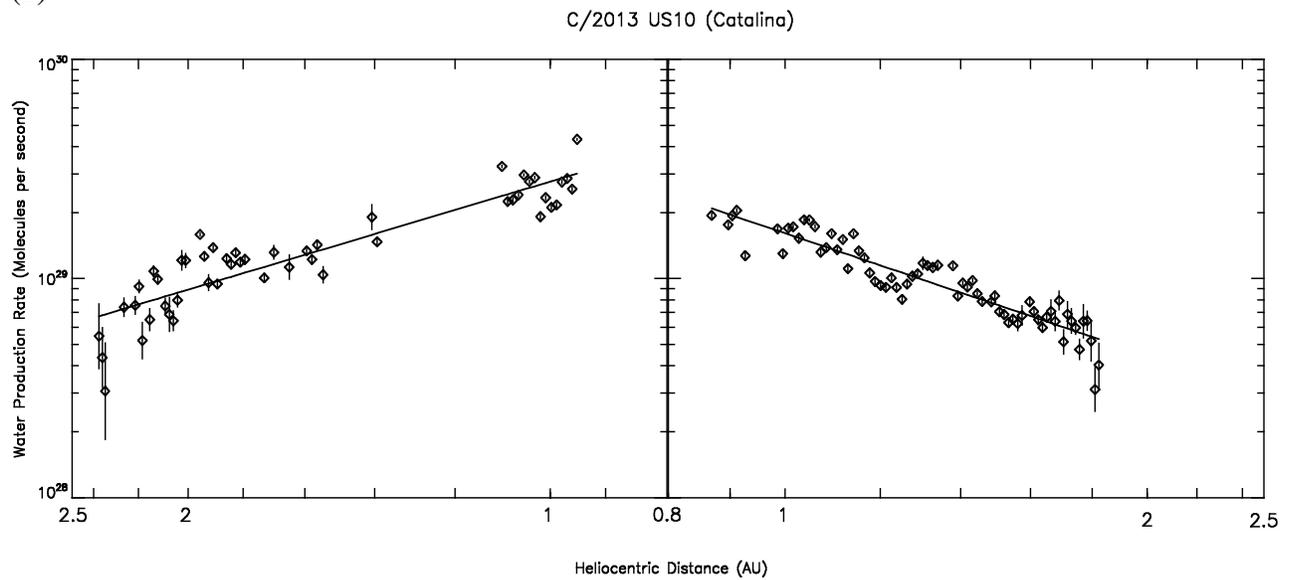

**Fig. 2.** Water production rate in comet C/2013 US10 (Catalina). (a) Variation over time in days from perihelion. (b) Variation with heliocentric distance. The error bars correspond to the 1-sigma stochastic errors from noise in the data and the model fitting. The lines give the best-fit power-law variation with form given in Table 2.

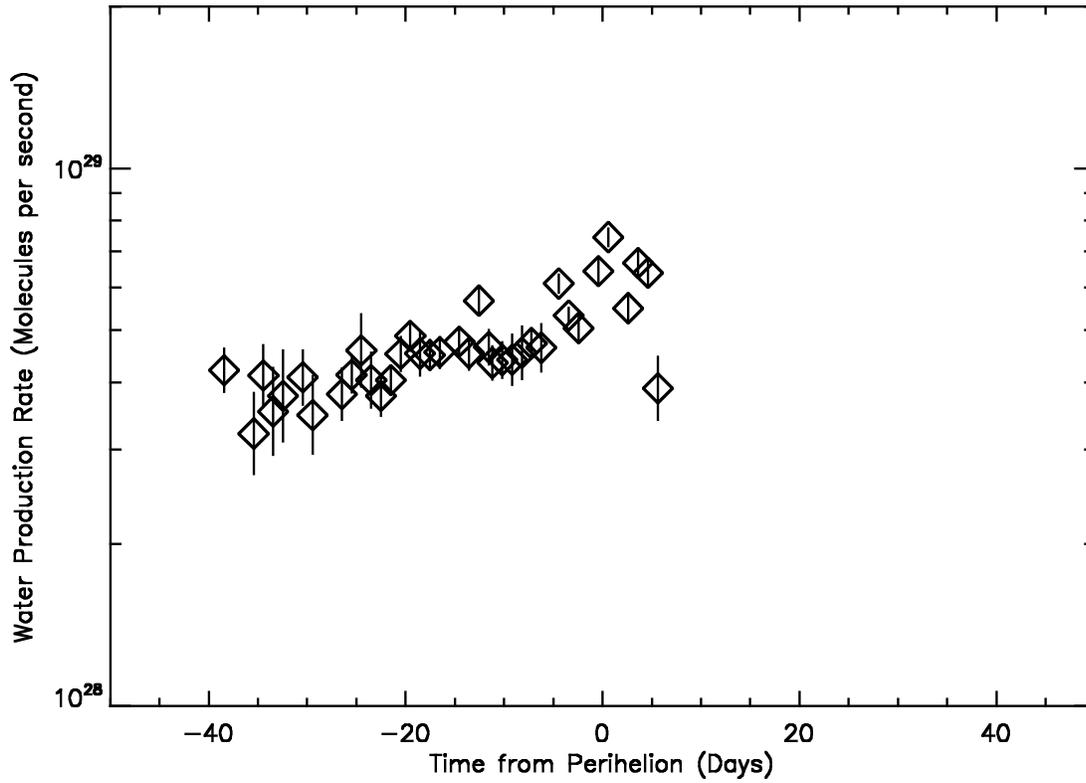

**Fig. 3.** Water production rate in comet C/2013 V5 (Oukaimeden) over time in days from perihelion. The error bars correspond to the 1-sigma stochastic errors from noise in the data and the model fitting.

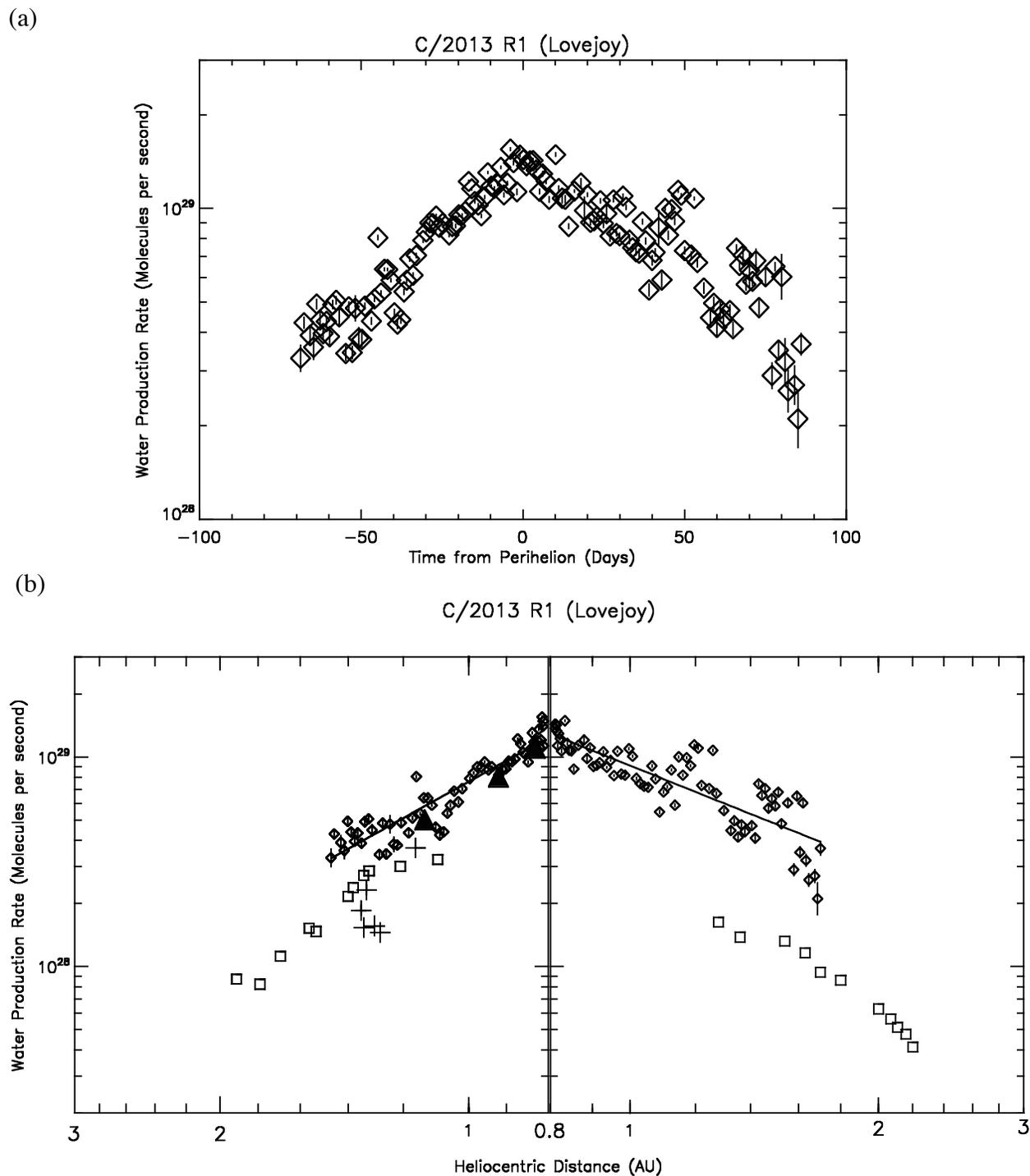

**Fig. 4.** Water production rate in comet C/2013 R1 (Lovejoy). (a) Variation over time in days from perihelion. (b) Variation with heliocentric distance. The error bars correspond to the 1-sigma stochastic errors from noise in the data and the model fitting. The lines give the best-fit power-law variation with form given in Table 2. The pluses give the production rates from Paganini et al. (2014), the filled triangles from Biver et al. (2014) and the squares from Opitom et al. (2015). The diamonds give the values from the SWAN measurements.

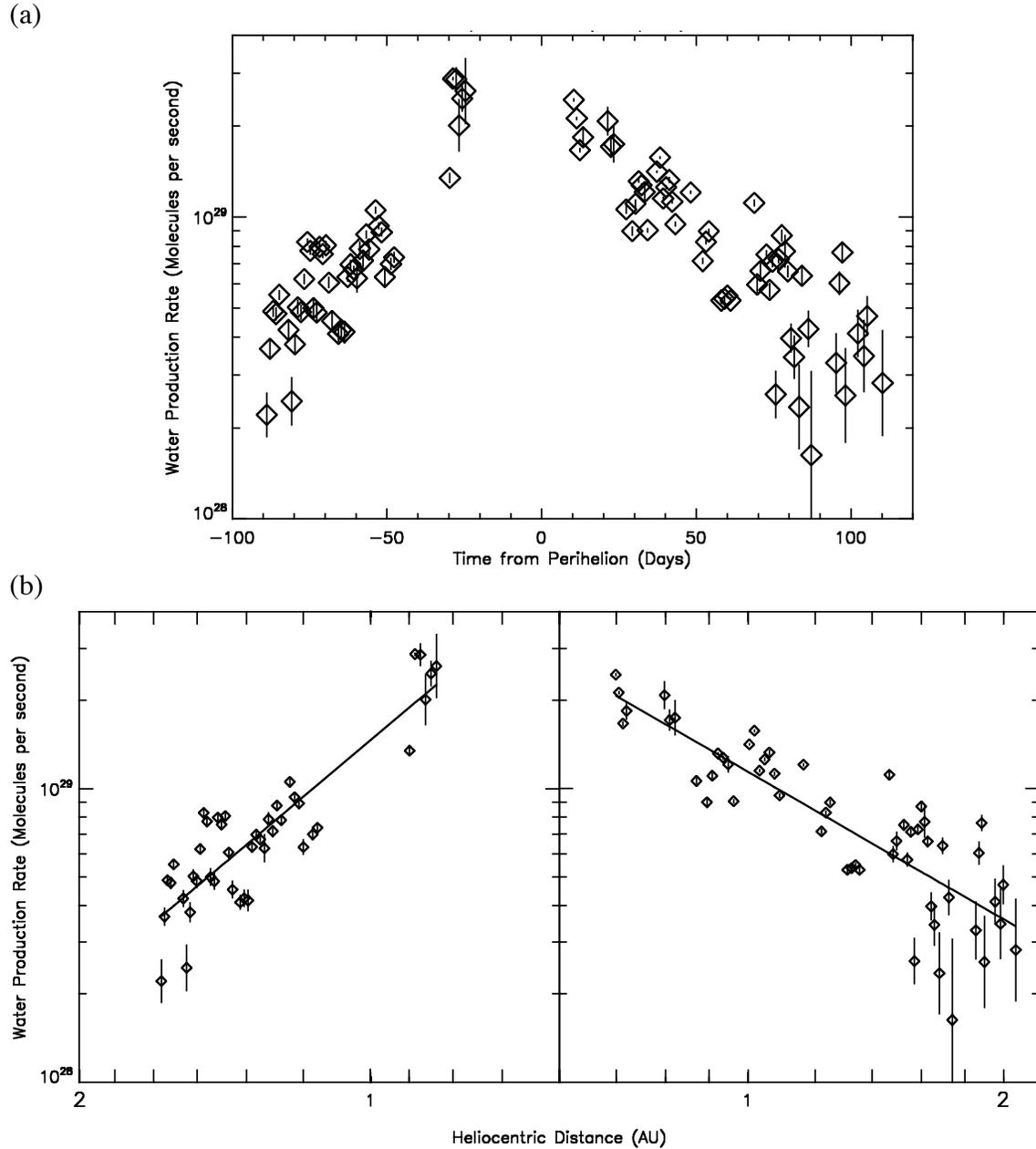

**Fig. 5.** Water production rate in comet C/2014 E2 (Jacques). (a) Variation over time in days from perihelion. (b) Variation with heliocentric distance. The error bars correspond to the 1-sigma stochastic errors from noise in the data and the model fitting. The lines give the best-fit power-law variation with form given in Table 2.

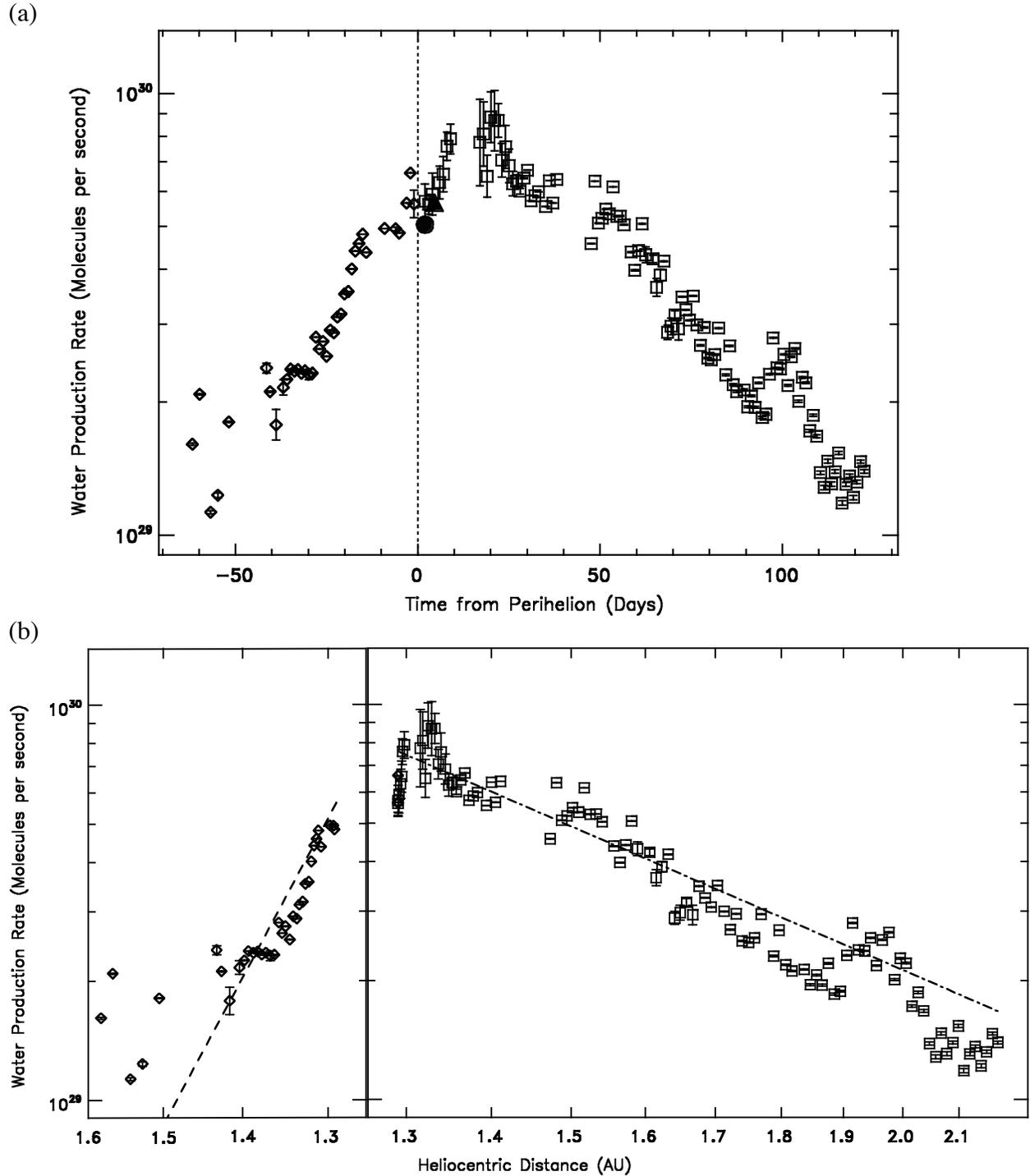

**Fig. 6.** Water production rate in comet C/2014 Q2 (Lovejoy). (a) Variation over time in days from perihelion. (b) Variation with heliocentric distance. The error bars correspond to the 1-sigma stochastic errors from noise in the data and the model fitting. The lines give the best-fit power-law variation with form given in Table 2. The filled circle in (a) shows the result from Faggi et al. (2016) and the filled triangle shows the result from Paganini et al. (2017).

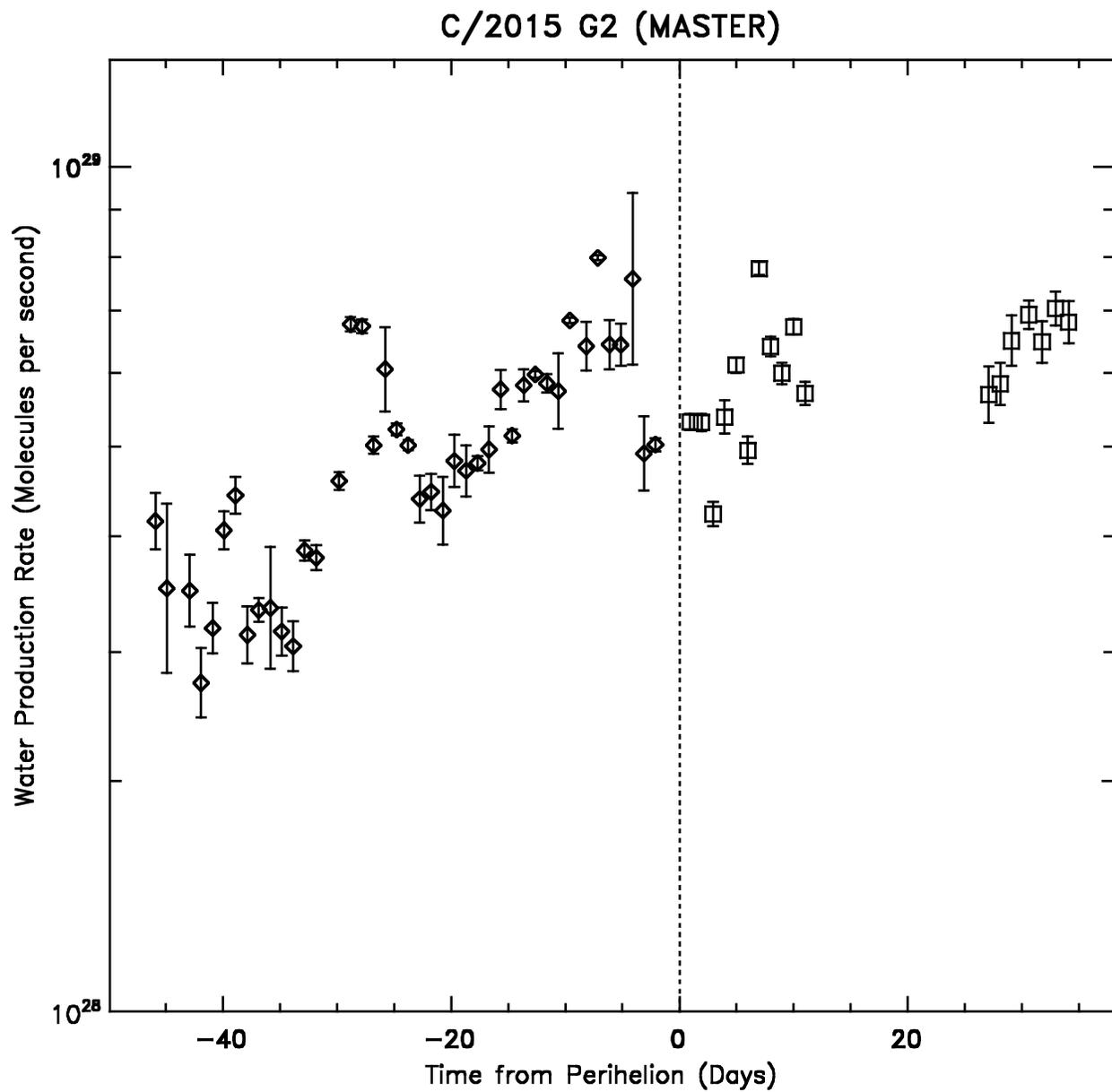

**Fig. 7.** Water production rate in comet C/2015 G2 (MASTER) over time in days from perihelion. The error bars correspond to the 1-sigma stochastic errors from noise in the data and the model fitting. The diamonds give the pre-perihelion values and the squares post-perihelion.

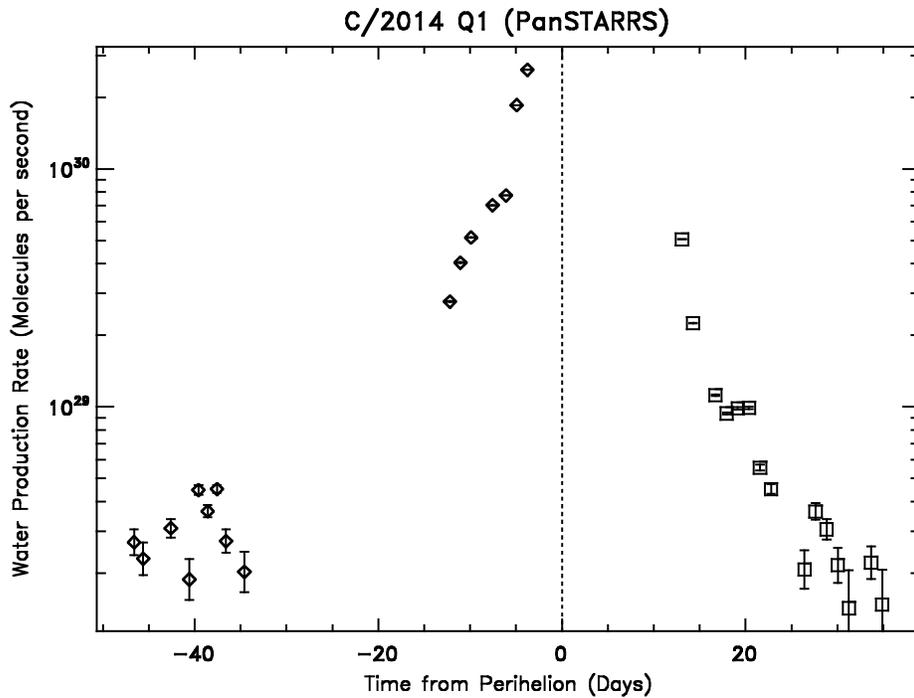

**Fig. 8.** Water production rate in comet C/2014 Q1 (PanSTARRS) over time in days from perihelion. The error bars correspond to the 1-sigma stochastic errors from noise in the data and the model fitting. The diamonds give the pre-perihelion values and the squares post-perihelion.

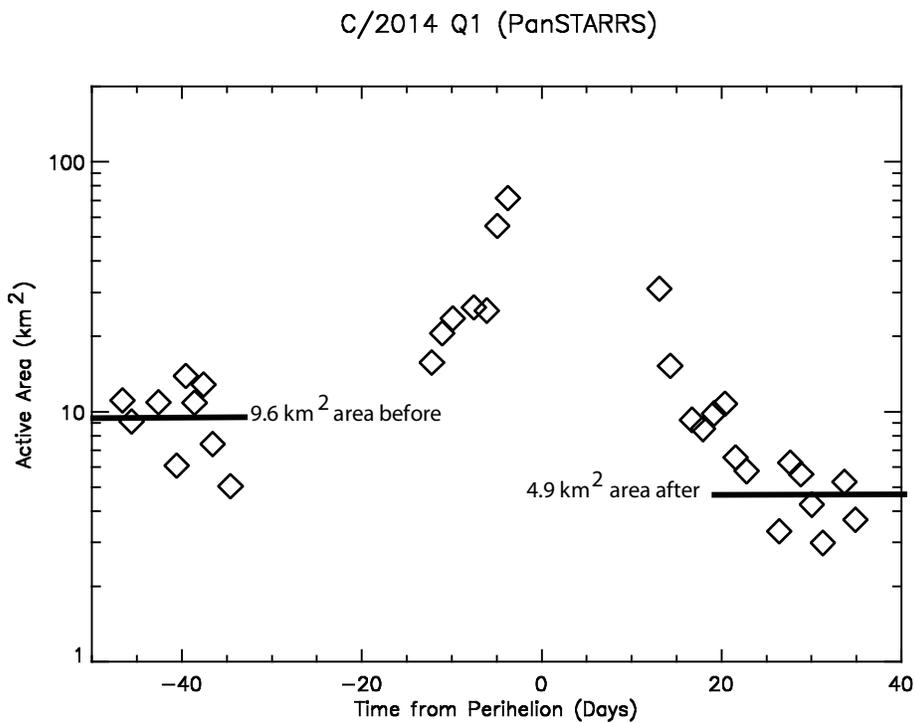

**Fig. 9.** Active area (km$^{-2}$) in comet C/2014 Q1 (PanSTARRS) over time in days from perihelion. The horizontal line on the left shows the average active area before the enhanced perihelion peak and the horizontal line on the right shows the average active area after.

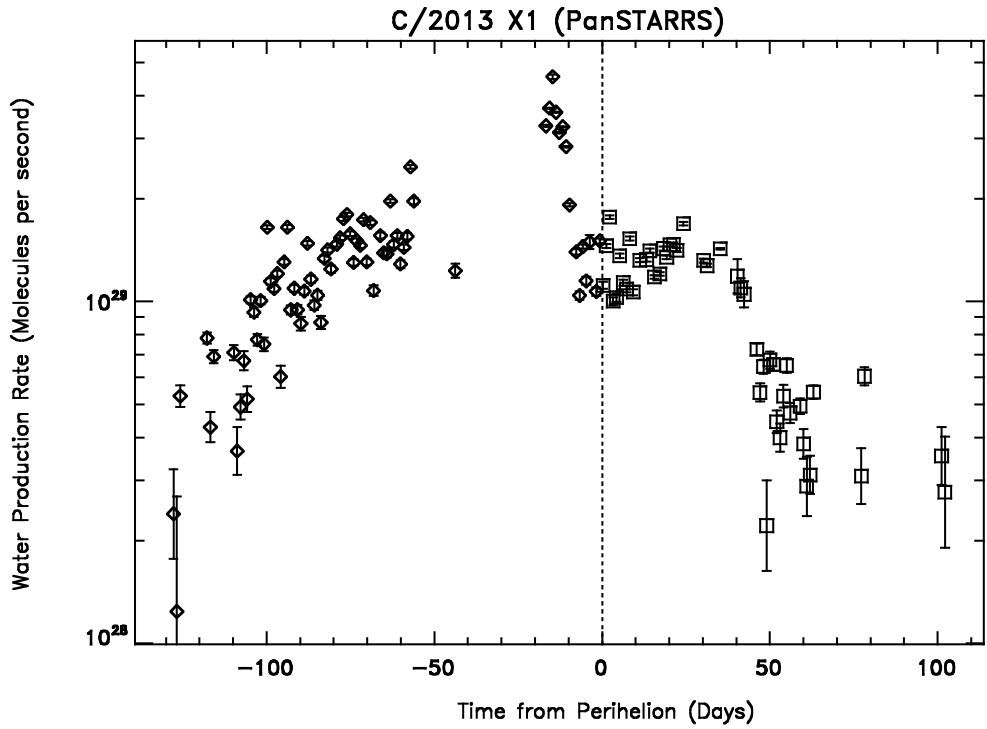
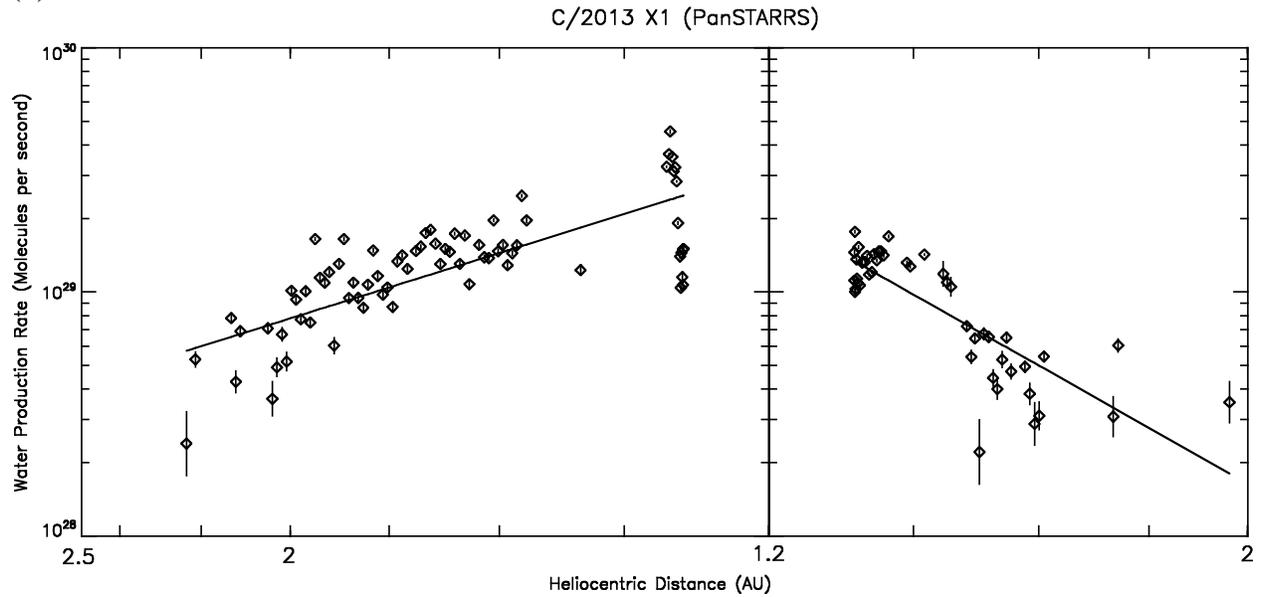

**Fig. 10.** Water production rate in comet C/2013 X1 (PanSTARRS). (a)Variation over time in days from perihelion. (b) Variation with heliocentric distance. The error bars correspond to the 1-sigma stochastic errors from noise in the data and the model fitting. The lines give the best-fit power-law variation with form given in Table 2.

**Supplemental Tables**

Table S1. SOHO/SWAN Observations of C/2012 K1 (PanSTARRS) and Water Production Rates

| ΔT (days) | r (AU) | Δ (AU) | g ($s^{-1}$) | Q ($10^{27}$ molecules $s^{-1}$) | δQ ($10^{27}$ molecules $s^{-1}$) |
|---|---|---|---|---|---|
| -126.088 | 2.197 | 1.515 | 0.002053 | 78.53 | 2.89 |
| -125.088 | 2.185 | 1.509 | 0.002075 | 96.81 | 2.09 |
| -124.088 | 2.173 | 1.503 | 0.002081 | 104.70 | 2.07 |
| -117.065 | 2.090 | 1.481 | 0.002255 | 97.04 | 2.35 |
| -116.065 | 2.078 | 1.481 | 0.002227 | 131.40 | 1.56 |
| -115.056 | 2.066 | 1.481 | 0.002145 | 133.90 | 1.40 |
| -114.056 | 2.054 | 1.482 | 0.002103 | 130.30 | 1.79 |
| -113.056 | 2.042 | 1.483 | 0.002095 | 74.51 | 2.87 |
| -112.056 | 2.031 | 1.485 | 0.002075 | 59.03 | 3.06 |
| -111.037 | 2.018 | 1.487 | 0.002046 | 76.98 | 3.00 |
| -110.037 | 2.007 | 1.490 | 0.002054 | 115.20 | 2.25 |
| -109.037 | 1.995 | 1.493 | 0.002061 | 85.29 | 2.58 |
| -108.027 | 1.983 | 1.497 | 0.002041 | 120.20 | 2.07 |
| -107.027 | 1.971 | 1.502 | 0.002028 | 104.90 | 2.41 |
| -105.026 | 1.947 | 1.512 | 0.001999 | 74.11 | 3.35 |
| -104.009 | 1.935 | 1.518 | 0.002005 | 104.20 | 2.59 |
| -103.009 | 1.924 | 1.524 | 0.001976 | 101.50 | 2.59 |
| -102.009 | 1.912 | 1.531 | 0.001977 | 89.76 | 3.24 |
| -100.997 | 1.900 | 1.538 | 0.002009 | 89.63 | 2.61 |
| -99.996 | 1.888 | 1.545 | 0.002029 | 87.39 | 3.31 |
| -98.997 | 1.876 | 1.553 | 0.002041 | 56.72 | 5.37 |
| -97.996 | 1.865 | 1.561 | 0.002076 | 85.57 | 3.80 |
| -96.982 | 1.853 | 1.570 | 0.002123 | 82.42 | 4.00 |
| -95.982 | 1.841 | 1.579 | 0.002148 | 85.34 | 2.91 |
| -94.982 | 1.829 | 1.588 | 0.002160 | 89.59 | 2.48 |
| -91.967 | 1.794 | 1.617 | 0.002159 | 107.90 | 2.57 |
| -90.966 | 1.782 | 1.627 | 0.002135 | 133.80 | 1.81 |
| -89.953 | 1.771 | 1.638 | 0.002109 | 113.60 | 2.69 |
| -88.954 | 1.759 | 1.648 | 0.002090 | 124.70 | 2.01 |
| -87.954 | 1.747 | 1.659 | 0.002026 | 175.90 | 1.92 |
| -86.954 | 1.736 | 1.670 | 0.002006 | 114.00 | 3.34 |
| -85.937 | 1.724 | 1.681 | 0.002006 | 100.50 | 2.33 |
| -84.937 | 1.712 | 1.692 | 0.002011 | 74.67 | 3.65 |
| -83.937 | 1.701 | 1.703 | 0.001966 | 91.79 | 3.76 |
| -82.925 | 1.689 | 1.714 | 0.001940 | 99.24 | 3.15 |
| -81.925 | 1.678 | 1.726 | 0.001887 | 105.40 | 3.27 |
| -80.925 | 1.666 | 1.737 | 0.001879 | 82.00 | 4.22 |
| -79.926 | 1.655 | 1.749 | 0.001870 | 89.14 | 4.08 |
| -78.908 | 1.643 | 1.760 | 0.001853 | 99.92 | 3.70 |
| -77.907 | 1.632 | 1.772 | 0.001855 | 94.43 | 4.06 |
| -76.907 | 1.621 | 1.783 | 0.001862 | 84.89 | 5.16 |
| -75.907 | 1.610 | 1.795 | 0.001881 | 90.36 | 4.17 |
| -73.897 | 1.587 | 1.818 | 0.002006 | 85.32 | 4.74 |
| -72.898 | 1.576 | 1.829 | 0.002078 | 85.00 | 3.61 |
| -71.898 | 1.565 | 1.840 | 0.002131 | 71.56 | 4.41 |
| -69.878 | 1.542 | 1.863 | 0.002183 | 94.67 | 3.31 |
| -68.878 | 1.531 | 1.874 | 0.002193 | 68.16 | 4.21 |
| -67.878 | 1.520 | 1.885 | 0.002202 | 77.32 | 3.88 |
| -66.878 | 1.509 | 1.896 | 0.002178 | 66.48 | 4.17 |
| -65.878 | 1.499 | 1.906 | 0.002149 | 87.60 | 3.52 |
| -64.676 | 1.486 | 1.919 | 0.002094 | 120.90 | 1.96 |
| -63.676 | 1.475 | 1.929 | 0.002000 | 144.60 | 2.11 |
| -62.676 | 1.464 | 1.940 | 0.001989 | 139.70 | 2.06 |
| -61.676 | 1.454 | 1.950 | 0.001977 | 194.10 | 1.76 |
| -60.698 | 1.443 | 1.959 | 0.001860 | 167.70 | 2.01 |
| -59.698 | 1.433 | 1.969 | 0.001834 | 182.80 | 2.05 |
| -58.698 | 1.422 | 1.978 | 0.001815 | 196.20 | 2.29 |
| -56.698 | 1.402 | 1.997 | 0.001792 | 196.60 | 2.02 |
| -55.698 | 1.392 | 2.005 | 0.001776 | 169.60 | 2.33 |
| -53.704 | 1.372 | 2.022 | 0.001801 | 161.40 | 2.30 |
| -52.704 | 1.362 | 2.030 | 0.001811 | 161.90 | 2.23 |
| -51.704 | 1.352 | 2.038 | 0.001849 | 122.60 | 3.48 |
| -50.704 | 1.342 | 2.045 | 0.002052 | 116.50 | 2.21 |
| -49.704 | 1.332 | 2.053 | 0.002081 | 156.00 | 2.32 |
| -47.727 | 1.314 | 2.066 | 0.002294 | 108.90 | 1.96 |
| -46.727 | 1.304 | 2.072 | 0.002327 | 106.90 | 2.08 |
| -45.728 | 1.295 | 2.079 | 0.002335 | 91.96 | 2.30 |
| -2.794 | 1.056 | 1.976 | 0.001836 | 192.50 | 2.51 |
| -1.817 | 1.055 | 1.964 | 0.001878 | 167.00 | 2.62 |
| -0.818 | 1.055 | 1.952 | 0.001888 | 182.60 | 2.32 |
| 1.183 | 1.055 | 1.927 | 0.001897 | 151.70 | 1.91 |
| 2.183 | 1.055 | 1.914 | 0.001892 | 171.50 | 2.51 |
| 4.180 | 1.057 | 1.886 | 0.001883 | 151.30 | 2.92 |
| 6.180 | 1.060 | 1.857 | 0.001859 | 159.40 | 2.58 |
| 8.159 | 1.063 | 1.826 | 0.001792 | 137.30 | 2.68 |
| 9.159 | 1.066 | 1.811 | 0.001811 | 187.40 | 2.52 |
| 10.159 | 1.068 | 1.794 | 0.001839 | 139.20 | 2.89 |
| 11.159 | 1.071 | 1.778 | 0.001866 | 156.60 | 2.46 |
| 12.159 | 1.074 | 1.761 | 0.001874 | 155.90 | 3.64 |
| 13.153 | 1.077 | 1.744 | 0.001883 | 140.90 | 2.77 |
| 14.153 | 1.081 | 1.727 | 0.001883 | 163.60 | 2.56 |
| 15.153 | 1.085 | 1.709 | 0.001883 | 143.60 | 2.51 |
| 17.153 | 1.093 | 1.673 | 0.001890 | 149.20 | 3.80 |
| 18.133 | 1.097 | 1.655 | 0.001880 | 122.50 | 2.43 |
| 19.133 | 1.102 | 1.636 | 0.001878 | 120.50 | 2.57 |
| 20.133 | 1.107 | 1.617 | 0.001853 | 112.40 | 2.56 |
| 21.126 | 1.112 | 1.598 | 0.001862 | 130.40 | 2.38 |
| 22.126 | 1.117 | 1.579 | 0.001868 | 153.00 | 2.00 |
| 23.126 | 1.123 | 1.560 | 0.001860 | 134.50 | 2.13 |
| 24.126 | 1.128 | 1.540 | 0.001870 | 147.30 | 1.97 |
| 25.105 | 1.134 | 1.521 | 0.001873 | 114.10 | 2.22 |
| 34.383 | 1.198 | 1.335 | 0.002018 | 104.60 | 1.95 |
| 35.383 | 1.206 | 1.316 | 0.002033 | 118.90 | 1.07 |
| 36.383 | 1.214 | 1.296 | 0.002033 | 138.70 | 0.93 |
| 37.403 | 1.222 | 1.276 | 0.002004 | 139.30 | 0.92 |
| 38.403 | 1.231 | 1.257 | 0.001999 | 142.50 | 0.93 |
| 39.412 | 1.239 | 1.238 | 0.001920 | 118.80 | 6.79 |
| 41.431 | 1.256 | 1.200 | 0.001878 | 113.20 | 1.38 |
| 43.441 | 1.274 | 1.164 | 0.001875 | 107.50 | 1.89 |
| 45.459 | 1.292 | 1.129 | 0.001880 | 92.66 | 1.62 |
| 46.459 | 1.302 | 1.113 | 0.001878 | 92.63 | 1.56 |
| 47.471 | 1.311 | 1.097 | 0.001892 | 88.61 | 1.23 |
| 51.500 | 1.350 | 1.039 | 0.002138 | 69.77 | 1.28 |
| 52.516 | 1.360 | 1.026 | 0.002068 | 63.40 | 1.25 |
| 53.529 | 1.370 | 1.014 | 0.002096 | 75.38 | 1.11 |
| 62.618 | 1.464 | 0.952 | 0.001939 | 66.20 | 0.97 |
| 66.648 | 1.507 | 0.955 | 0.001956 | 100.50 | 0.66 |
| 67.649 | 1.518 | 0.959 | 0.001948 | 70.27 | 0.73 |
| 68.650 | 1.529 | 0.964 | 0.001945 | 55.91 | 0.96 |
| 69.651 | 1.540 | 0.970 | 0.001935 | 52.65 | 1.09 |
| 70.651 | 1.551 | 0.977 | 0.001943 | 48.53 | 1.05 |
| 71.810 | 1.564 | 0.987 | 0.002044 | 49.00 | 1.15 |
| 73.837 | 1.586 | 1.008 | 0.002175 | 42.54 | 1.14 |
| 74.837 | 1.597 | 1.020 | 0.002183 | 46.23 | 0.96 |
| 76.842 | 1.620 | 1.047 | 0.002187 | 46.35 | 0.91 |
| 77.864 | 1.632 | 1.063 | 0.002187 | 38.47 | 1.43 |
| 78.864 | 1.643 | 1.079 | 0.002175 | 50.98 | 0.88 |
| 79.871 | 1.654 | 1.096 | 0.002178 | 55.55 | 1.04 |
| 80.871 | 1.666 | 1.114 | 0.002169 | 54.03 | 1.65 |
| 81.892 | 1.678 | 1.133 | 0.002128 | 51.69 | 1.28 |
| 82.892 | 1.689 | 1.152 | 0.002076 | 48.03 | 1.81 |
| 83.901 | 1.701 | 1.172 | 0.002110 | 54.78 | 1.59 |
| 84.901 | 1.712 | 1.193 | 0.002158 | 36.38 | 2.21 |

| ΔT | r | Δ | g | Q | δQ |
|---|---|---|---|---|---|
| 85.920 | 1.724 | 1.215 | 0.002167 | 44.97 | 1.99 |
| 86.920 | 1.735 | 1.237 | 0.002162 | 61.65 | 1.02 |
| 87.931 | 1.747 | 1.260 | 0.002155 | 50.26 | 1.72 |
| 88.931 | 1.759 | 1.284 | 0.002109 | 51.01 | 1.40 |
| 90.948 | 1.782 | 1.332 | 0.002049 | 39.68 | 2.08 |
| 92.960 | 1.806 | 1.382 | 0.001983 | 25.92 | 4.15 |
| 93.960 | 1.817 | 1.408 | 0.001953 | 40.24 | 3.81 |
| 94.960 | 1.829 | 1.434 | 0.001932 | 36.78 | 2.85 |
| 95.976 | 1.841 | 1.461 | 0.001951 | 43.37 | 2.67 |
| 96.976 | 1.853 | 1.487 | 0.002105 | 31.04 | 3.32 |
| 97.989 | 1.865 | 1.515 | 0.002165 | 31.28 | 3.15 |
| 98.990 | 1.876 | 1.542 | 0.002192 | 34.20 | 2.65 |
| 99.989 | 1.888 | 1.569 | 0.002228 | 28.72 | 2.13 |
| 100.990 | 1.900 | 1.597 | 0.002227 | 28.58 | 3.54 |
| 102.004 | 1.912 | 1.625 | 0.002237 | 41.31 | 2.32 |

ΔT: Time from perihelion on 27.7 Aug 2014 UT in days
r : Heliocentric distance (AU)
Δ: Geocentric distance (AU)
g: Solar Lyman-α g-factor (photons s$^{-1}$) at 1 AU
Q: Daily-Average Water production rates (molecules s$^{-1}$) from the TRM
δQ: internal 1-sigma uncertainties

Table S2. SOHO/SWAN Observations of C/2013 US10 (Catalina) and Water Production Rates

| ΔT (days) | r (AU) | Δ (AU) | g ($s^{-1}$) | Q ($10^{27}$ molecules $s^{-1}$) | δQ ($10^{27}$ molecules $s^{-1}$) |
|---|---|---|---|---|---|
| -137.270 | 2.375 | 1.793 | 0.001572 | 54.62 | 22.07 |
| -136.103 | 2.360 | 1.762 | 0.001837 | 43.59 | 16.22 |
| -134.956 | 2.346 | 1.731 | 0.001934 | 30.63 | 20.19 |
| -128.390 | 2.261 | 1.567 | 0.002192 | 73.97 | 7.48 |
| -124.625 | 2.213 | 1.480 | 0.002129 | 75.54 | 7.28 |
| -123.425 | 2.197 | 1.453 | 0.002099 | 92.04 | 6.27 |
| -122.356 | 2.183 | 1.430 | 0.002086 | 52.13 | 10.83 |
| -119.953 | 2.152 | 1.380 | 0.002038 | 64.94 | 7.95 |
| -118.744 | 2.136 | 1.357 | 0.002007 | 108.00 | 6.06 |
| -117.546 | 2.120 | 1.334 | 0.001972 | 99.28 | 5.11 |
| -115.122 | 2.089 | 1.290 | 0.001906 | 74.93 | 6.41 |
| -113.914 | 2.073 | 1.270 | 0.001880 | 68.61 | 13.35 |
| -112.708 | 2.057 | 1.251 | 0.001840 | 64.09 | 7.00 |
| -111.499 | 2.041 | 1.233 | 0.001841 | 79.63 | 5.71 |
| -110.291 | 2.025 | 1.216 | 0.001851 | 121.20 | 12.85 |
| -109.082 | 2.009 | 1.200 | 0.001941 | 121.00 | 9.00 |
| -104.950 | 1.955 | 1.154 | 0.002065 | 158.90 | 2.15 |
| -103.742 | 1.939 | 1.143 | 0.002065 | 126.20 | 5.42 |
| -102.533 | 1.923 | 1.133 | 0.002070 | 95.80 | 8.21 |
| -101.324 | 1.907 | 1.125 | 0.002067 | 138.50 | 2.84 |
| -100.116 | 1.891 | 1.118 | 0.002063 | 94.46 | 4.37 |
| -97.698 | 1.858 | 1.108 | 0.002010 | 123.40 | 3.91 |
| -96.490 | 1.842 | 1.104 | 0.001992 | 116.30 | 4.03 |
| -95.281 | 1.826 | 1.102 | 0.001975 | 131.20 | 4.02 |
| -94.072 | 1.810 | 1.102 | 0.001966 | 118.90 | 4.17 |
| -92.864 | 1.794 | 1.102 | 0.001930 | 122.30 | 4.55 |
| -88.025 | 1.729 | 1.115 | 0.001859 | 100.60 | 3.75 |
| -85.598 | 1.697 | 1.128 | 0.001918 | 131.70 | 9.55 |
| -81.970 | 1.648 | 1.155 | 0.001955 | 112.70 | 15.17 |
| -77.935 | 1.594 | 1.193 | 0.001968 | 133.70 | 6.60 |
| -76.727 | 1.578 | 1.205 | 0.001954 | 122.20 | 4.94 |
| -75.532 | 1.562 | 1.219 | 0.001959 | 142.60 | 5.72 |
| -74.324 | 1.545 | 1.233 | 0.001949 | 104.10 | 9.00 |
| -63.913 | 1.407 | 1.367 | 0.001834 | 190.70 | 26.81 |
| -62.894 | 1.393 | 1.381 | 0.001856 | 147.20 | 4.33 |
| -39.350 | 1.097 | 1.682 | 0.001705 | 325.20 | 4.27 |
| -38.345 | 1.085 | 1.693 | 0.001719 | 224.70 | 6.14 |
| -37.345 | 1.074 | 1.702 | 0.001742 | 228.60 | 5.23 |
| -36.345 | 1.063 | 1.712 | 0.001751 | 241.10 | 4.49 |
| -35.345 | 1.052 | 1.721 | 0.001785 | 296.90 | 2.74 |
| -34.345 | 1.041 | 1.730 | 0.001815 | 277.20 | 4.17 |
| -33.345 | 1.030 | 1.738 | 0.001849 | 288.90 | 3.75 |
| -32.345 | 1.019 | 1.747 | 0.001895 | 191.60 | 5.23 |
| -31.345 | 1.009 | 1.754 | 0.001957 | 233.80 | 4.19 |
| -30.321 | 0.998 | 1.762 | 0.001984 | 211.10 | 4.75 |
| -29.320 | 0.988 | 1.768 | 0.002002 | 217.00 | 4.89 |
| -28.320 | 0.978 | 1.775 | 0.001984 | 275.70 | 4.12 |
| -27.320 | 0.968 | 1.781 | 0.001961 | 286.50 | 4.18 |
| -26.320 | 0.959 | 1.786 | 0.001930 | 256.30 | 4.79 |
| -25.320 | 0.950 | 1.791 | 0.001898 | 432.20 | 3.80 |
| 14.739 | 0.869 | 1.526 | 0.001389 | 194.10 | 5.85 |
| 18.903 | 0.897 | 1.449 | 0.001675 | 175.90 | 4.38 |
| 19.904 | 0.904 | 1.430 | 0.001716 | 193.40 | 3.57 |
| 20.961 | 0.912 | 1.409 | 0.001765 | 204.80 | 3.67 |
| 22.732 | 0.927 | 1.373 | 0.001844 | 127.10 | 4.92 |
| 29.081 | 0.986 | 1.240 | 0.001812 | 168.70 | 3.92 |
| 30.065 | 0.996 | 1.218 | 0.001799 | 129.80 | 3.57 |
| 31.065 | 1.006 | 1.197 | 0.001815 | 169.90 | 2.79 |
| 32.057 | 1.016 | 1.175 | 0.001823 | 172.40 | 2.88 |
| 33.057 | 1.027 | 1.153 | 0.001830 | 152.80 | 3.37 |
| 34.037 | 1.037 | 1.132 | 0.001828 | 185.70 | 8.22 |
| 35.037 | 1.048 | 1.110 | 0.001808 | 185.10 | 2.33 |
| 36.037 | 1.059 | 1.088 | 0.001783 | 172.70 | 2.88 |
| 37.037 | 1.071 | 1.067 | 0.001785 | 132.10 | 3.21 |
| 38.027 | 1.082 | 1.046 | 0.001768 | 138.30 | 2.77 |
| 39.027 | 1.093 | 1.024 | 0.001750 | 160.40 | 2.56 |
| 40.009 | 1.105 | 1.004 | 0.001734 | 135.40 | 3.35 |
| 41.009 | 1.117 | 0.983 | 0.001721 | 150.90 | 2.61 |
| 41.998 | 1.128 | 0.963 | 0.001725 | 111.00 | 3.52 |
| 42.998 | 1.140 | 0.943 | 0.001717 | 160.20 | 2.48 |
| 43.981 | 1.152 | 0.923 | 0.001713 | 133.90 | 2.79 |
| 44.981 | 1.164 | 0.904 | 0.001726 | 124.40 | 2.73 |
| 45.969 | 1.176 | 0.886 | 0.001762 | 106.10 | 3.24 |
| 46.953 | 1.188 | 0.868 | 0.001814 | 96.99 | 4.44 |
| 47.953 | 1.201 | 0.851 | 0.001845 | 92.97 | 2.91 |
| 48.940 | 1.213 | 0.835 | 0.001892 | 91.23 | 2.55 |
| 49.924 | 1.226 | 0.819 | 0.001897 | 100.70 | 2.18 |
| 50.924 | 1.238 | 0.804 | 0.001902 | 91.12 | 2.26 |
| 51.910 | 1.251 | 0.790 | 0.001881 | 80.42 | 3.31 |
| 52.896 | 1.263 | 0.778 | 0.001867 | 94.31 | 4.46 |

| ΔT | r | Δ | | | |
|---|---|---|---|---|---|
| 53.881 | 1.276 | 0.766 | 0.001840 | 102.50 | 1.93 |
| 54.881 | 1.289 | 0.755 | 0.001833 | 105.10 | 2.09 |
| 55.868 | 1.302 | 0.746 | 0.001827 | 117.50 | 7.83 |
| 56.852 | 1.314 | 0.738 | 0.001825 | 114.70 | 4.68 |
| 57.840 | 1.327 | 0.731 | 0.001823 | 112.40 | 1.80 |
| 58.840 | 1.340 | 0.726 | 0.001827 | 115.00 | 1.72 |
| 61.813 | 1.379 | 0.719 | 0.001810 | 114.50 | 2.25 |
| 62.792 | 1.392 | 0.720 | 0.001793 | 83.31 | 2.08 |
| 63.792 | 1.405 | 0.722 | 0.001807 | 95.38 | 1.79 |
| 64.786 | 1.418 | 0.726 | 0.001805 | 91.51 | 4.40 |
| 65.786 | 1.432 | 0.732 | 0.001800 | 98.07 | 1.68 |
| 66.787 | 1.445 | 0.739 | 0.001792 | 85.44 | 1.83 |
| 67.761 | 1.458 | 0.748 | 0.001791 | 78.61 | 1.81 |
| 69.760 | 1.484 | 0.769 | 0.001767 | 78.35 | 1.94 |
| 70.458 | 1.494 | 0.778 | 0.001792 | 83.43 | 1.96 |
| 71.459 | 1.507 | 0.792 | 0.001830 | 70.56 | 2.09 |
| 72.459 | 1.521 | 0.807 | 0.001874 | 68.51 | 2.10 |
| 73.460 | 1.534 | 0.823 | 0.001919 | 62.97 | 1.71 |
| 74.456 | 1.547 | 0.841 | 0.001927 | 65.18 | 2.96 |
| 75.456 | 1.561 | 0.859 | 0.001946 | 62.37 | 4.80 |
| 76.432 | 1.574 | 0.878 | 0.001925 | 67.86 | 7.14 |
| 78.160 | 1.597 | 0.914 | 0.001911 | 78.38 | 2.44 |
| 79.153 | 1.610 | 0.936 | 0.001918 | 70.81 | 2.40 |
| 80.153 | 1.624 | 0.959 | 0.001902 | 64.89 | 3.20 |
| 81.132 | 1.637 | 0.982 | 0.001867 | 59.80 | 3.13 |
| 82.132 | 1.650 | 1.006 | 0.001855 | 66.73 | 2.82 |
| 83.132 | 1.664 | 1.031 | 0.001849 | 70.83 | 9.14 |
| 84.124 | 1.677 | 1.056 | 0.001850 | 63.77 | 6.47 |
| 85.124 | 1.690 | 1.081 | 0.001823 | 79.33 | 8.16 |
| 86.124 | 1.704 | 1.108 | 0.001817 | 51.42 | 6.89 |
| 87.105 | 1.717 | 1.134 | 0.001793 | 68.66 | 9.81 |
| 88.105 | 1.730 | 1.161 | 0.001799 | 63.98 | 8.79 |
| 89.105 | 1.744 | 1.189 | 0.001803 | 59.62 | 4.12 |
| 90.094 | 1.757 | 1.216 | 0.001800 | 47.50 | 5.23 |
| 91.094 | 1.770 | 1.244 | 0.001791 | 63.86 | 12.19 |
| 92.077 | 1.783 | 1.272 | 0.001803 | 64.22 | 6.63 |
| 93.077 | 1.797 | 1.301 | 0.001796 | 51.95 | 12.27 |
| 94.077 | 1.810 | 1.330 | 0.001777 | 31.20 | 8.13 |
| 95.077 | 1.823 | 1.359 | 0.001773 | 40.32 | 10.33 |

ΔT: Time from perihelion on 15.7 Nov 2015 UT in days
r : Heliocentric distance (AU)
Δ: Geocentric distance (AU)

g: Solar Lyman-α g-factor (photons s$^{-1}$) at 1 AU
Q: Daily-Average Water production rates (molecules s$^{-1}$) from the TRM
δQ: internal 1-sigma uncertainties

Table S3. SOHO/SWAN Observations of C/2013 V5 (Oukaimeden) and Water Production Rates

| ΔT (days) | r (AU) | Δ (AU) | g (s$^{-1}$) | Q (10$^{27}$ molecules s$^{-1}$) | δQ (10$^{27}$ molecules s$^{-1}$) |
|---|---|---|---|---|---|
| -38.435 | 1.010 | 1.230 | 0.002013 | 42.12 | 4.06 |
| -35.435 | 0.965 | 1.122 | 0.002109 | 32.13 | 6.07 |
| -34.444 | 0.951 | 1.087 | 0.002161 | 41.18 | 5.75 |
| -33.444 | 0.936 | 1.051 | 0.002156 | 35.31 | 7.22 |
| -32.444 | 0.922 | 1.014 | 0.002155 | 37.72 | 8.08 |
| -30.464 | 0.894 | 0.943 | 0.002171 | 40.85 | 5.02 |
| -29.464 | 0.880 | 0.907 | 0.002159 | 34.78 | 6.29 |
| -26.472 | 0.839 | 0.802 | 0.002193 | 38.01 | 4.39 |
| -25.493 | 0.826 | 0.769 | 0.002067 | 41.28 | 3.16 |
| -24.493 | 0.813 | 0.735 | 0.002022 | 45.84 | 7.71 |
| -23.493 | 0.800 | 0.702 | 0.002004 | 40.35 | 5.04 |
| -22.501 | 0.788 | 0.671 | 0.001995 | 37.75 | 3.36 |
| -21.501 | 0.775 | 0.640 | 0.002028 | 40.36 | 2.54 |
| -20.522 | 0.764 | 0.612 | 0.002057 | 45.15 | 3.39 |
| -19.522 | 0.752 | 0.585 | 0.002045 | 48.81 | 3.03 |
| -18.529 | 0.741 | 0.560 | 0.002042 | 45.23 | 4.50 |
| -17.529 | 0.730 | 0.537 | 0.002038 | 44.98 | 2.92 |
| -16.552 | 0.719 | 0.517 | 0.002021 | 45.48 | 3.10 |
| -14.553 | 0.700 | 0.488 | 0.001951 | 47.62 | 2.94 |
| -13.553 | 0.690 | 0.479 | 0.001916 | 45.35 | 3.36 |
| -12.554 | 0.682 | 0.474 | 0.001884 | 56.73 | 3.16 |
| -11.554 | 0.673 | 0.474 | 0.001855 | 46.49 | 3.51 |
| -11.172 | 0.670 | 0.475 | 0.001858 | 43.46 | 3.24 |
| -10.171 | 0.663 | 0.481 | 0.001860 | 44.01 | 3.47 |
| -9.173 | 0.656 | 0.491 | 0.001835 | 44.07 | 4.99 |
| -8.199 | 0.650 | 0.505 | 0.001840 | 45.36 | 5.38 |
| -7.202 | 0.645 | 0.523 | 0.001825 | 47.30 | 2.75 |
| -6.227 | 0.640 | 0.544 | 0.001815 | 46.38 | 4.93 |
| -4.456 | 0.633 | 0.589 | 0.001832 | 61.02 | 2.34 |
| -3.432 | 0.630 | 0.619 | 0.001850 | 53.26 | 1.81 |
| -2.432 | 0.628 | 0.649 | 0.001869 | 50.47 | 2.77 |
| -0.404 | 0.626 | 0.716 | 0.001872 | 64.37 | 3.39 |
| 0.596 | 0.626 | 0.750 | 0.001869 | 74.47 | 2.84 |
| 2.602 | 0.628 | 0.823 | 0.001846 | 54.94 | 3.33 |
| 3.620 | 0.630 | 0.859 | 0.001839 | 66.71 | 3.40 |
| 4.620 | 0.633 | 0.896 | 0.001848 | 63.88 | 3.73 |
| 5.622 | 0.637 | 0.933 | 0.001878 | 38.98 | 5.57 |

| ΔT | r | Δ | g | Q | δQ |
|---|---|---|---|---|---|
| 6.622 | 0.642 | 0.970 | 0.001898 | 40.55 | 6.02 |
| 8.634 | 0.653 | 1.044 | 0.001878 | 40.39 | 6.52 |
| 9.633 | 0.659 | 1.081 | 0.001872 | 36.78 | 6.12 |
| 10.633 | 0.666 | 1.117 | 0.001898 | 32.90 | 6.10 |

ΔT: Time from perihelion on 28.2 Sep 2014 UT in days
r : Heliocentric distance (AU)
Δ: Geocentric distance (AU)
g: Solar Lyman-α g-factor (photons s$^{-1}$) at 1 AU
Q: Daily-Average Water production rates (molecules s$^{-1}$) from the TRM
δQ: internal 1-sigma uncertainties

Table S4. SOHO/SWAN Observations of C/2013 R1 (Lovejoy) and Water Production Rates

| ΔT (days) | r (AU) | Δ (AU) | g ($s^{-1}$) | Q ($10^{27}$ molecules $s^{-1}$) | δQ ($10^{27}$ molecules $s^{-1}$) |
|---|---|---|---|---|---|
| -69.730 | 1.482 | 1.030 | 0.001760 | 15.83 | 6.94 |
| -68.730 | 1.468 | 1.006 | 0.001707 | 33.01 | 3.33 |
| -67.730 | 1.455 | 0.982 | 0.001689 | 42.92 | 1.77 |
| -66.743 | 1.442 | 0.959 | 0.001688 | 12.09 | 9.60 |
| -65.743 | 1.428 | 0.935 | 0.001688 | 39.04 | 2.99 |
| -64.743 | 1.415 | 0.911 | 0.001751 | 35.75 | 3.31 |
| -63.743 | 1.401 | 0.888 | 0.001799 | 49.29 | 1.78 |
| -62.743 | 1.388 | 0.865 | 0.001899 | 43.78 | 2.38 |
| -61.758 | 1.375 | 0.842 | 0.001942 | 39.59 | 1.69 |
| -60.758 | 1.362 | 0.819 | 0.001918 | 43.32 | 1.54 |
| -59.758 | 1.349 | 0.796 | 0.001899 | 38.72 | 2.09 |
| -58.772 | 1.336 | 0.774 | 0.001883 | 49.09 | 1.50 |
| -57.771 | 1.322 | 0.752 | 0.001913 | 50.64 | 1.41 |
| -56.772 | 1.309 | 0.729 | 0.001859 | 44.75 | 2.76 |
| -54.787 | 1.283 | 0.686 | 0.001850 | 34.18 | 1.80 |
| -53.800 | 1.271 | 0.665 | 0.001849 | 48.36 | 1.60 |
| -52.800 | 1.258 | 0.644 | 0.001900 | 34.44 | 1.61 |
| -51.800 | 1.245 | 0.623 | 0.001885 | 47.74 | 4.65 |
| -50.812 | 1.232 | 0.603 | 0.001884 | 38.32 | 3.08 |
| -49.858 | 1.220 | 0.584 | 0.001943 | 37.93 | 2.01 |
| -48.857 | 1.207 | 0.565 | 0.001911 | 48.55 | 1.01 |
| -46.857 | 1.182 | 0.528 | 0.001866 | 43.46 | 1.08 |
| -45.858 | 1.170 | 0.510 | 0.001850 | 51.17 | 0.86 |
| -44.837 | 1.157 | 0.493 | 0.001738 | 80.59 | 1.17 |
| -43.851 | 1.145 | 0.478 | 0.001797 | 53.57 | 0.75 |
| -42.838 | 1.133 | 0.463 | 0.001840 | 63.79 | 0.62 |
| -41.831 | 1.120 | 0.449 | 0.001819 | 63.61 | 0.60 |
| -40.818 | 1.108 | 0.436 | 0.001858 | 58.78 | 0.63 |
| -39.804 | 1.096 | 0.424 | 0.001886 | 46.14 | 1.29 |
| -38.789 | 1.084 | 0.414 | 0.001913 | 42.42 | 1.38 |
| -37.761 | 1.072 | 0.405 | 0.001949 | 43.93 | 1.21 |
| -36.746 | 1.061 | 0.398 | 0.001885 | 53.92 | 1.07 |
| -35.996 | 1.052 | 0.394 | 0.001904 | 58.72 | 0.97 |
| -34.981 | 1.041 | 0.389 | 0.001839 | 68.81 | 1.09 |
| -33.967 | 1.029 | 0.387 | 0.001775 | 61.00 | 1.08 |
| -32.952 | 1.018 | 0.386 | 0.001753 | 70.56 | 1.04 |
| -30.910 | 0.997 | 0.390 | 0.001721 | 78.92 | 1.06 |
| -29.894 | 0.986 | 0.395 | 0.001670 | 84.07 | 0.82 |
| -28.882 | 0.976 | 0.401 | 0.001647 | 89.85 | 0.92 |
| -27.864 | 0.965 | 0.409 | 0.001642 | 89.45 | 0.99 |
| -26.864 | 0.956 | 0.418 | 0.001613 | 94.64 | 0.94 |
| -25.854 | 0.946 | 0.428 | 0.001640 | 86.99 | 1.00 |
| -24.835 | 0.937 | 0.440 | 0.001651 | 89.64 | 0.83 |
| -22.835 | 0.919 | 0.467 | 0.001683 | 82.26 | 1.04 |
| -21.826 | 0.910 | 0.482 | 0.001664 | 87.50 | 1.00 |
| -20.827 | 0.902 | 0.498 | 0.001649 | 87.77 | 1.08 |
| -19.827 | 0.894 | 0.514 | 0.001631 | 95.34 | 1.00 |
| -18.827 | 0.886 | 0.531 | 0.001655 | 95.06 | 0.98 |
| -17.827 | 0.879 | 0.549 | 0.001637 | 97.92 | 1.06 |
| -16.827 | 0.872 | 0.567 | 0.001639 | 122.10 | 0.94 |
| -15.828 | 0.865 | 0.585 | 0.001689 | 115.70 | 1.06 |
| -14.828 | 0.859 | 0.604 | 0.001715 | 105.20 | 1.16 |
| -13.828 | 0.853 | 0.623 | 0.001737 | 102.40 | 1.08 |
| -12.828 | 0.847 | 0.642 | 0.001763 | 94.61 | 1.18 |
| -11.828 | 0.842 | 0.662 | 0.001757 | 108.80 | 1.13 |
| -10.832 | 0.837 | 0.682 | 0.001752 | 130.40 | 1.03 |
| -9.832 | 0.833 | 0.702 | 0.001769 | 118.00 | 1.06 |
| -8.832 | 0.829 | 0.722 | 0.001763 | 116.10 | 1.10 |
| -7.832 | 0.825 | 0.742 | 0.001765 | 119.60 | 1.23 |
| -6.832 | 0.822 | 0.762 | 0.001697 | 135.60 | 1.13 |
| -5.832 | 0.819 | 0.782 | 0.001692 | 110.50 | 1.49 |
| -4.857 | 0.817 | 0.801 | 0.001598 | 120.50 | 1.66 |
| -3.857 | 0.815 | 0.822 | 0.001548 | 155.10 | 1.37 |
| -2.857 | 0.813 | 0.842 | 0.001501 | 140.20 | 1.83 |
| -1.857 | 0.812 | 0.861 | 0.001497 | 113.10 | 2.24 |
| -0.860 | 0.812 | 0.881 | 0.001511 | 148.20 | 1.57 |
| 0.140 | 0.812 | 0.901 | 0.001509 | 143.50 | 1.80 |
| 1.140 | 0.812 | 0.920 | 0.001465 | 137.10 | 1.93 |
| 2.140 | 0.813 | 0.940 | 0.001526 | 142.20 | 1.75 |
| 3.140 | 0.814 | 0.959 | 0.001541 | 142.50 | 2.03 |
| 4.114 | 0.815 | 0.977 | 0.001510 | 132.90 | 1.92 |
| 5.114 | 0.817 | 0.996 | 0.001528 | 113.30 | 1.91 |
| 6.114 | 0.820 | 1.015 | 0.001513 | 129.10 | 2.19 |
| 7.114 | 0.823 | 1.033 | 0.001514 | 122.30 | 1.90 |
| 8.115 | 0.826 | 1.051 | 0.001595 | 106.90 | 2.16 |
| 10.111 | 0.834 | 1.086 | 0.001549 | 149.00 | 1.81 |
| 11.111 | 0.839 | 1.103 | 0.001579 | 116.00 | 1.97 |
| 12.111 | 0.844 | 1.120 | 0.001640 | 107.90 | 4.02 |
| 13.111 | 0.849 | 1.137 | 0.001694 | 107.00 | 2.84 |
| 14.111 | 0.855 | 1.153 | 0.001737 | 87.55 | 1.51 |
| 15.953 | 0.866 | 1.182 | 0.001793 | 113.70 | 1.11 |
| 17.953 | 0.880 | 1.213 | 0.001689 | 120.60 | 3.78 |
| 18.953 | 0.887 | 1.228 | 0.001691 | 98.35 | 10.45 |
| 19.914 | 0.895 | 1.242 | 0.001651 | 110.60 | 1.44 |
| 20.914 | 0.903 | 1.256 | 0.001674 | 90.16 | 1.56 |
| 21.918 | 0.911 | 1.270 | 0.001623 | 91.10 | 1.82 |
| 22.918 | 0.919 | 1.283 | 0.001616 | 94.05 | 1.79 |
| 23.918 | 0.928 | 1.297 | 0.001652 | 105.90 | 1.70 |
| 24.918 | 0.937 | 1.310 | 0.001643 | 89.32 | 2.45 |
| 25.918 | 0.947 | 1.322 | 0.001685 | 96.26 | 7.16 |
| 26.918 | 0.956 | 1.334 | 0.001748 | 81.53 | 3.20 |
| 27.942 | 0.966 | 1.346 | 0.001809 | 106.50 | 1.44 |
| 28.942 | 0.976 | 1.358 | 0.001778 | 82.99 | 1.47 |
| 29.942 | 0.986 | 1.369 | 0.001758 | 81.76 | 1.56 |
| 30.942 | 0.997 | 1.380 | 0.001685 | 109.50 | 1.73 |
| 31.942 | 1.007 | 1.390 | 0.001710 | 100.80 | 1.70 |
| 32.942 | 1.018 | 1.401 | 0.001755 | 79.03 | 1.90 |
| 33.942 | 1.029 | 1.410 | 0.001749 | 74.79 | 1.64 |
| 34.948 | 1.040 | 1.420 | 0.001704 | 72.52 | 1.75 |
| 35.948 | 1.052 | 1.429 | 0.001737 | 71.76 | 2.18 |
| 36.948 | 1.063 | 1.438 | 0.001732 | 90.68 | 1.56 |
| 37.948 | 1.074 | 1.447 | 0.001745 | 78.60 | 1.76 |
| 38.948 | 1.086 | 1.455 | 0.001732 | 54.66 | 2.41 |
| 39.948 | 1.098 | 1.463 | 0.001736 | 68.03 | 2.39 |
| 40.948 | 1.110 | 1.470 | 0.001713 | 72.19 | 1.96 |
| 41.970 | 1.122 | 1.478 | 0.001700 | 86.79 | 11.55 |
| 42.970 | 1.134 | 1.485 | 0.001585 | 58.86 | 3.14 |
| 43.970 | 1.146 | 1.492 | 0.001613 | 100.20 | 1.45 |
| 44.970 | 1.159 | 1.498 | 0.001610 | 81.98 | 2.55 |
| 45.970 | 1.171 | 1.504 | 0.001585 | 99.30 | 2.20 |
| 46.970 | 1.184 | 1.510 | 0.001603 | 90.95 | 2.04 |
| 47.970 | 1.196 | 1.515 | 0.001571 | 114.30 | 1.44 |
| 48.970 | 1.209 | 1.521 | 0.001648 | 110.20 | 1.37 |
| 49.970 | 1.221 | 1.526 | 0.001717 | 73.21 | 1.80 |
| 51.978 | 1.247 | 1.535 | 0.001790 | 70.77 | 1.66 |
| 52.978 | 1.260 | 1.539 | 0.001777 | 107.60 | 1.18 |
| 53.978 | 1.273 | 1.543 | 0.001777 | 66.95 | 2.06 |
| 55.978 | 1.299 | 1.550 | 0.001838 | 55.46 | 2.19 |
| 56.978 | 1.312 | 1.553 | 0.001891 | 24.11 | 4.50 |
| 57.978 | 1.325 | 1.556 | 0.001897 | 44.56 | 2.59 |
| 58.978 | 1.338 | 1.558 | 0.001957 | 49.55 | 2.15 |
| 59.998 | 1.352 | 1.561 | 0.001937 | 41.43 | 2.95 |
| 60.998 | 1.365 | 1.563 | 0.001931 | 47.43 | 2.32 |

| ΔT | r | Δ | g | Q | δQ |
|---|---|---|---|---|---|
| 61.998 | 1.378 | 1.564 | 0.001930 | 44.05 | 2.67 |
| 63.998 | 1.405 | 1.568 | 0.001905 | 46.93 | 2.01 |
| 64.998 | 1.418 | 1.569 | 0.001884 | 41.02 | 2.06 |
| 65.998 | 1.432 | 1.570 | 0.001842 | 74.45 | 1.73 |
| 66.998 | 1.445 | 1.570 | 0.001743 | 65.56 | 1.99 |
| 67.998 | 1.458 | 1.571 | 0.001747 | 70.66 | 5.60 |
| 68.998 | 1.472 | 1.571 | 0.001753 | 56.96 | 2.42 |
| 69.998 | 1.485 | 1.572 | 0.001743 | 63.26 | 8.48 |
| 71.008 | 1.499 | 1.571 | 0.001725 | 58.32 | 1.66 |
| 72.008 | 1.512 | 1.571 | 0.001781 | 67.84 | 6.00 |
| 73.008 | 1.526 | 1.571 | 0.001785 | 48.00 | 2.54 |
| 75.008 | 1.553 | 1.570 | 0.001771 | 60.29 | 4.08 |
| 77.009 | 1.579 | 1.568 | 0.001847 | 29.00 | 2.90 |
| 78.009 | 1.593 | 1.566 | 0.001840 | 64.99 | 1.95 |
| 79.009 | 1.606 | 1.565 | 0.001863 | 35.03 | 2.20 |
| 80.025 | 1.620 | 1.563 | 0.001925 | 60.28 | 10.85 |
| 81.025 | 1.634 | 1.562 | 0.001908 | 32.12 | 5.93 |
| 82.025 | 1.647 | 1.560 | 0.001940 | 25.89 | 4.29 |
| 83.025 | 1.661 | 1.558 | 0.001940 | 58.86 | 1.87 |
| 84.025 | 1.674 | 1.556 | 0.001949 | 27.05 | 4.12 |
| 85.025 | 1.688 | 1.554 | 0.001938 | 21.05 | 4.93 |
| 86.025 | 1.701 | 1.551 | 0.001928 | 36.62 | 3.01 |
| 87.025 | 1.714 | 1.549 | 0.001881 | 8.47 | 12.28 |
| 91.039 | 1.768 | 1.539 | 0.001745 | 54.90 | 3.00 |

ΔT: Time from perihelion on 22.7 Dec 2013 UT in days
r : Heliocentric distance (AU)
Δ: Geocentric distance (AU)
g: Solar Lyman-α g-factor (photons s$^{-1}$) at 1 AU
Q: Daily-Average Water production rates (molecules s$^{-1}$) from the TRM
δQ: internal 1-sigma uncertainties

Table S5. SOHO/SWAN Observations of C/2014 E2 (Jacques) and Water Production Rates

| ΔT (days) | r (AU) | Δ (AU) | g ($s^{-1}$) | Q ($10^{27}$ molecules $s^{-1}$) | δQ ($10^{27}$ molecules $s^{-1}$) |
|---|---|---|---|---|---|
| -88.844 | 1.763 | 0.972 | 0.002045 | 22.12 | 4.00 |
| -87.844 | 1.748 | 0.972 | 0.002026 | 36.63 | 2.53 |
| -86.822 | 1.734 | 0.973 | 0.002033 | 48.78 | 1.83 |
| -85.843 | 1.719 | 0.974 | 0.002025 | 47.77 | 2.17 |
| -84.822 | 1.705 | 0.976 | 0.002049 | 55.22 | 1.85 |
| -81.823 | 1.661 | 0.988 | 0.001930 | 42.25 | 2.71 |
| -80.813 | 1.646 | 0.994 | 0.001914 | 24.53 | 4.80 |
| -79.813 | 1.631 | 1.001 | 0.001891 | 37.91 | 2.90 |
| -78.812 | 1.617 | 1.008 | 0.001859 | 50.31 | 2.60 |
| -77.812 | 1.602 | 1.016 | 0.001804 | 48.48 | 2.61 |
| -76.812 | 1.587 | 1.024 | 0.001808 | 62.30 | 2.09 |
| -75.795 | 1.572 | 1.034 | 0.001731 | 82.73 | 1.81 |
| -74.795 | 1.558 | 1.044 | 0.001699 | 77.31 | 2.09 |
| -73.795 | 1.543 | 1.054 | 0.001732 | 49.97 | 3.33 |
| -72.796 | 1.528 | 1.065 | 0.001684 | 48.33 | 2.97 |
| -71.782 | 1.513 | 1.076 | 0.001694 | 79.70 | 1.97 |
| -70.782 | 1.498 | 1.088 | 0.001715 | 75.42 | 1.93 |
| -69.782 | 1.483 | 1.100 | 0.001741 | 80.70 | 2.15 |
| -68.782 | 1.469 | 1.113 | 0.001762 | 60.67 | 2.37 |
| -67.782 | 1.454 | 1.126 | 0.001851 | 45.29 | 3.03 |
| -65.768 | 1.424 | 1.153 | 0.001887 | 41.02 | 2.27 |
| -64.768 | 1.409 | 1.167 | 0.002036 | 42.27 | 2.76 |
| -63.752 | 1.394 | 1.181 | 0.001990 | 41.61 | 3.50 |
| -62.752 | 1.379 | 1.195 | 0.001946 | 63.33 | 2.00 |
| -61.752 | 1.364 | 1.209 | 0.001932 | 69.69 | 1.97 |
| -60.752 | 1.349 | 1.224 | 0.001925 | 67.16 | 1.31 |
| -59.751 | 1.334 | 1.238 | 0.001908 | 62.70 | 6.97 |
| -58.740 | 1.319 | 1.253 | 0.001852 | 78.61 | 4.13 |
| -57.751 | 1.304 | 1.268 | 0.001882 | 71.68 | 3.07 |
| -56.741 | 1.289 | 1.283 | 0.001856 | 87.54 | 2.26 |
| -55.741 | 1.274 | 1.298 | 0.001853 | 78.10 | 2.91 |
| -53.741 | 1.245 | 1.327 | 0.001870 | 105.40 | 2.27 |
| -52.722 | 1.229 | 1.342 | 0.001869 | 93.49 | 2.12 |
| -51.721 | 1.214 | 1.357 | 0.001858 | 89.01 | 2.63 |
| -50.721 | 1.200 | 1.371 | 0.001752 | 63.25 | 3.77 |
| -48.721 | 1.170 | 1.400 | 0.001808 | 69.89 | 2.88 |
| -47.721 | 1.155 | 1.414 | 0.001872 | 73.58 | 2.84 |
| -29.686 | 0.900 | 1.626 | 0.001714 | 134.70 | 4.58 |
| -28.686 | 0.887 | 1.634 | 0.001770 | 287.40 | 2.20 |
| -27.686 | 0.874 | 1.642 | 0.001721 | 286.00 | 25.63 |
| -26.686 | 0.862 | 1.650 | 0.001722 | 201.00 | 43.38 |
| -25.686 | 0.849 | 1.657 | 0.001685 | 246.60 | 24.63 |
| -24.686 | 0.837 | 1.663 | 0.001696 | 261.40 | 73.46 |
| 10.346 | 0.699 | 1.455 | 0.001837 | 244.30 | 2.03 |
| 11.347 | 0.705 | 1.436 | 0.001946 | 212.20 | 1.77 |
| 12.347 | 0.713 | 1.417 | 0.001870 | 166.60 | 2.26 |
| 13.347 | 0.720 | 1.397 | 0.001852 | 183.70 | 12.20 |
| 21.338 | 0.798 | 1.222 | 0.001504 | 207.70 | 22.89 |
| 22.335 | 0.809 | 1.199 | 0.001518 | 171.00 | 14.01 |
| 23.335 | 0.821 | 1.175 | 0.001526 | 174.20 | 25.07 |
| 27.319 | 0.870 | 1.080 | 0.001581 | 106.10 | 3.09 |
| 29.311 | 0.895 | 1.031 | 0.001659 | 89.84 | 2.83 |
| 30.311 | 0.908 | 1.007 | 0.001785 | 110.60 | 3.34 |
| 31.310 | 0.922 | 0.982 | 0.001850 | 131.60 | 0.97 |
| 32.291 | 0.935 | 0.958 | 0.001929 | 127.50 | 1.37 |
| 33.291 | 0.948 | 0.934 | 0.001958 | 120.90 | 7.49 |
| 34.281 | 0.962 | 0.910 | 0.001946 | 90.49 | 1.12 |
| 37.262 | 1.004 | 0.840 | 0.001773 | 141.40 | 0.78 |
| 38.263 | 1.018 | 0.817 | 0.001753 | 157.40 | 0.89 |
| 39.252 | 1.032 | 0.794 | 0.001686 | 115.10 | 1.04 |
| 40.234 | 1.046 | 0.772 | 0.001626 | 125.50 | 1.20 |
| 41.223 | 1.060 | 0.751 | 0.001570 | 132.70 | 2.29 |
| 42.223 | 1.075 | 0.730 | 0.001547 | 112.40 | 1.11 |
| 43.206 | 1.089 | 0.710 | 0.001541 | 94.77 | 1.19 |
| 48.150 | 1.162 | 0.624 | 0.001420 | 120.70 | 0.92 |
| 52.093 | 1.220 | 0.578 | 0.001612 | 71.48 | 1.29 |
| 53.077 | 1.235 | 0.570 | 0.001703 | 82.78 | 1.09 |
| 54.066 | 1.249 | 0.565 | 0.001643 | 89.74 | 1.29 |
| 58.039 | 1.309 | 0.562 | 0.001803 | 52.89 | 1.08 |
| 59.040 | 1.324 | 0.566 | 0.001803 | 53.57 | 0.91 |
| 60.015 | 1.338 | 0.573 | 0.001863 | 54.90 | 0.79 |
| 61.015 | 1.353 | 0.581 | 0.001855 | 52.90 | 1.02 |
| 68.681 | 1.467 | 0.701 | 0.001533 | 111.40 | 2.32 |
| 69.658 | 1.482 | 0.722 | 0.001533 | 59.77 | 3.53 |
| 70.658 | 1.496 | 0.745 | 0.001533 | 66.29 | 4.69 |
| 72.651 | 1.526 | 0.792 | 0.001533 | 75.20 | 2.03 |
| 73.630 | 1.540 | 0.817 | 0.001533 | 57.31 | 2.98 |
| 74.630 | 1.555 | 0.843 | 0.001533 | 71.19 | 2.75 |
| 75.622 | 1.570 | 0.869 | 0.001533 | 25.85 | 4.96 |
| 76.621 | 1.585 | 0.896 | 0.001533 | 72.58 | 1.90 |

| ΔT | r | Δ | g | Q | δQ |
|---|---|---|---|---|---|
| 77.602 | 1.599 | 0.923 | 0.001533 | 86.93 | 3.55 |
| 78.602 | 1.614 | 0.951 | 0.001533 | 77.01 | 9.80 |
| 79.592 | 1.628 | 0.979 | 0.001533 | 66.03 | 2.70 |
| 80.592 | 1.643 | 1.008 | 0.001533 | 39.71 | 4.42 |
| 81.592 | 1.657 | 1.038 | 0.001533 | 34.31 | 5.91 |
| 83.155 | 1.680 | 1.084 | 0.001533 | 23.49 | 8.76 |
| 84.156 | 1.695 | 1.114 | 0.001532 | 63.88 | 3.95 |
| 86.167 | 1.724 | 1.175 | 0.001532 | 42.59 | 6.10 |
| 87.167 | 1.739 | 1.206 | 0.001532 | 16.28 | 14.43 |
| 95.196 | 1.854 | 1.460 | 0.001519 | 32.86 | 8.20 |
| 96.196 | 1.869 | 1.492 | 0.001518 | 60.33 | 5.40 |
| 97.196 | 1.883 | 1.524 | 0.001518 | 76.36 | 4.60 |
| 98.196 | 1.897 | 1.556 | 0.001518 | 25.66 | 11.01 |
| 102.212 | 1.954 | 1.685 | 0.001518 | 41.16 | 7.89 |
| 104.212 | 1.982 | 1.750 | 0.001518 | 34.62 | 10.75 |
| 105.212 | 1.996 | 1.782 | 0.001518 | 47.01 | 7.50 |
| 110.226 | 2.066 | 1.943 | 0.001518 | 28.20 | 13.81 |

ΔT: Time from perihelion on 2.5 Jul 2013 UT in days
r : Heliocentric distance (AU)
Δ: Geocentric distance (AU)
g: Solar Lyman-α g-factor (photons s$^{-1}$) at 1 AU
Q: Daily-Average Water production rates (molecules s$^{-1}$) from the TRM
δQ: internal 1-sigma uncertainties

Table S6. SOHO/SWAN Observations of C/2014 Q2 (Lovejoy) and Water Production Rates

| ΔT (days) | r (AU) | Δ (AU) | g ($s^{-1}$) | Q ($10^{27}$ molecules $s^{-1}$) | δQ ($10^{27}$ molecules $s^{-1}$) |
|---|---|---|---|---|---|
| -61.833 | 1.583 | 1.079 | 0.001901 | 160.70 | 0.86 |
| -59.832 | 1.567 | 1.037 | 0.001898 | 208.40 | 0.78 |
| -56.832 | 1.543 | 0.974 | 0.002050 | 112.70 | 0.87 |
| -54.832 | 1.527 | 0.932 | 0.001450 | 123.20 | 1.52 |
| -51.853 | 1.505 | 0.871 | 0.002164 | 180.50 | 0.44 |
| -41.494 | 1.432 | 0.673 | 0.001891 | 239.00 | 6.37 |
| -40.494 | 1.426 | 0.656 | 0.001873 | 211.30 | 0.59 |
| -38.853 | 1.416 | 0.628 | 0.001422 | 177.90 | 14.73 |
| -36.881 | 1.404 | 0.598 | 0.001818 | 216.00 | 8.85 |
| -35.881 | 1.398 | 0.584 | 0.001823 | 225.20 | 0.50 |
| -34.883 | 1.393 | 0.570 | 0.001832 | 237.80 | 0.42 |
| -33.883 | 1.387 | 0.557 | 0.001829 | 235.30 | 0.39 |
| -32.909 | 1.382 | 0.546 | 0.001840 | 237.10 | 0.41 |
| -31.913 | 1.377 | 0.535 | 0.001872 | 232.70 | 0.36 |
| -30.937 | 1.372 | 0.525 | 0.001887 | 235.60 | 0.47 |
| -29.937 | 1.367 | 0.516 | 0.001934 | 231.20 | 6.60 |
| -28.942 | 1.362 | 0.508 | 0.001986 | 232.70 | 0.37 |
| -27.971 | 1.357 | 0.502 | 0.002013 | 280.70 | 0.31 |
| -26.971 | 1.353 | 0.496 | 0.002031 | 263.90 | 0.34 |
| -25.994 | 1.349 | 0.492 | 0.002033 | 274.70 | 0.37 |
| -25.000 | 1.344 | 0.489 | 0.002104 | 254.20 | 0.37 |
| -24.021 | 1.340 | 0.487 | 0.002010 | 291.40 | 0.36 |
| -23.029 | 1.336 | 0.487 | 0.001988 | 287.20 | 0.37 |
| -22.050 | 1.333 | 0.488 | 0.001953 | 311.60 | 0.34 |
| -21.079 | 1.329 | 0.490 | 0.001951 | 317.00 | 0.39 |
| -20.087 | 1.326 | 0.494 | 0.001909 | 351.80 | 0.28 |
| -19.107 | 1.322 | 0.498 | 0.001870 | 356.10 | 0.31 |
| -18.116 | 1.319 | 0.504 | 0.001831 | 400.70 | 0.35 |
| -17.133 | 1.316 | 0.511 | 0.001787 | 439.70 | 0.33 |
| -16.137 | 1.313 | 0.519 | 0.001765 | 457.30 | 0.34 |
| -15.153 | 1.311 | 0.529 | 0.001752 | 479.90 | 0.34 |
| -14.157 | 1.308 | 0.539 | 0.001734 | 436.70 | 0.53 |
| -9.113 | 1.298 | 0.603 | 0.001801 | 494.60 | 0.26 |
| -6.088 | 1.294 | 0.650 | 0.001865 | 493.70 | 4.54 |
| -5.071 | 1.293 | 0.667 | 0.001886 | 483.10 | 0.23 |
| -3.059 | 1.291 | 0.701 | 0.001952 | 564.60 | 0.21 |
| -2.059 | 1.291 | 0.718 | 0.001973 | 660.80 | 0.18 |
| -1.043 | 1.291 | 0.737 | 0.001977 | 562.10 | 41.82 |
| 1.970 | 1.291 | 0.792 | 0.001972 | 570.00 | 54.26 |
| 2.985 | 1.291 | 0.811 | 0.001958 | 561.30 | 29.55 |
| 3.985 | 1.292 | 0.829 | 0.001940 | 591.40 | 68.05 |
| 6.000 | 1.294 | 0.868 | 0.001885 | 629.10 | 56.08 |
| 7.000 | 1.295 | 0.887 | 0.001833 | 656.30 | 62.83 |
| 8.000 | 1.296 | 0.906 | 0.001789 | 759.10 | 58.71 |
| 9.014 | 1.298 | 0.925 | 0.001763 | 789.10 | 63.31 |
| 17.042 | 1.316 | 1.078 | 0.001839 | 774.50 | 195.80 |
| 18.042 | 1.319 | 1.097 | 0.001794 | 809.60 | 147.70 |
| 19.042 | 1.322 | 1.116 | 0.001811 | 649.20 | 75.18 |
| 20.042 | 1.325 | 1.134 | 0.001836 | 883.60 | 124.70 |
| 21.059 | 1.329 | 1.153 | 0.001834 | 867.70 | 146.80 |
| 22.059 | 1.333 | 1.172 | 0.001857 | 868.70 | 80.13 |
| 23.059 | 1.337 | 1.190 | 0.001871 | 705.90 | 63.99 |
| 24.059 | 1.340 | 1.209 | 0.001872 | 757.30 | 89.22 |
| 25.059 | 1.345 | 1.227 | 0.001880 | 686.10 | 62.88 |
| 26.070 | 1.349 | 1.245 | 0.001892 | 624.50 | 41.28 |
| 27.070 | 1.353 | 1.263 | 0.001886 | 633.10 | 20.39 |
| 28.070 | 1.358 | 1.281 | 0.001878 | 601.20 | 8.96 |
| 29.070 | 1.363 | 1.299 | 0.001861 | 643.30 | 2.61 |
| 30.070 | 1.368 | 1.316 | 0.001860 | 670.40 | 0.25 |
| 31.070 | 1.373 | 1.333 | 0.001840 | 571.50 | 0.34 |
| 32.088 | 1.378 | 1.351 | 0.001836 | 586.40 | 0.27 |
| 33.088 | 1.383 | 1.368 | 0.001812 | 598.50 | 0.28 |
| 35.088 | 1.394 | 1.402 | 0.001754 | 554.30 | 0.36 |
| 36.088 | 1.400 | 1.419 | 0.001726 | 635.00 | 0.35 |
| 37.088 | 1.405 | 1.435 | 0.001701 | 565.10 | 0.35 |
| 38.098 | 1.411 | 1.451 | 0.001696 | 638.20 | 0.35 |
| 47.572 | 1.473 | 1.598 | 0.001391 | 456.80 | 0.72 |
| 48.572 | 1.481 | 1.612 | 0.001391 | 633.50 | 0.77 |
| 49.572 | 1.488 | 1.627 | 0.001391 | 508.70 | 0.72 |
| 50.572 | 1.495 | 1.641 | 0.001391 | 521.50 | 0.85 |
| 51.572 | 1.502 | 1.655 | 0.001391 | 547.50 | 0.72 |
| 52.572 | 1.510 | 1.669 | 0.001400 | 533.20 | 0.95 |
| 53.564 | 1.517 | 1.682 | 0.001400 | 614.40 | 0.62 |
| 54.564 | 1.525 | 1.696 | 0.001400 | 527.00 | 0.83 |
| 55.564 | 1.533 | 1.709 | 0.001400 | 528.00 | 0.77 |
| 56.564 | 1.541 | 1.722 | 0.001400 | 504.30 | 0.90 |
| 58.564 | 1.557 | 1.748 | 0.001400 | 437.50 | 1.12 |
| 59.564 | 1.565 | 1.761 | 0.001408 | 397.70 | 1.07 |
| 60.564 | 1.573 | 1.773 | 0.001408 | 440.40 | 0.89 |
| 61.543 | 1.581 | 1.785 | 0.001408 | 506.80 | 0.81 |
| 62.543 | 1.589 | 1.797 | 0.001408 | 431.20 | 17.66 |
| 64.543 | 1.606 | 1.821 | 0.001408 | 422.20 | 8.24 |
| 65.543 | 1.615 | 1.833 | 0.001408 | 363.90 | 17.36 |
| 66.543 | 1.623 | 1.845 | 0.001408 | 388.00 | 11.86 |
| 67.543 | 1.632 | 1.856 | 0.001417 | 417.10 | 0.88 |
| 68.543 | 1.641 | 1.867 | 0.001881 | 287.70 | 10.94 |
| 69.543 | 1.649 | 1.878 | 0.001929 | 297.00 | 13.22 |
| 70.536 | 1.658 | 1.889 | 0.001971 | 315.30 | 6.79 |
| 71.537 | 1.667 | 1.900 | 0.002014 | 293.20 | 17.21 |
| 72.536 | 1.676 | 1.910 | 0.002045 | 346.50 | 0.67 |
| 73.537 | 1.685 | 1.921 | 0.002033 | 323.70 | 0.73 |
| 74.536 | 1.694 | 1.931 | 0.001996 | 306.70 | 0.70 |
| 75.537 | 1.703 | 1.941 | 0.001960 | 347.90 | 0.72 |
| 76.537 | 1.713 | 1.951 | 0.001923 | 299.10 | 0.80 |
| 77.537 | 1.722 | 1.961 | 0.001895 | 269.00 | 0.84 |
| 78.537 | 1.731 | 1.971 | 0.001879 | 295.00 | 0.85 |
| 79.514 | 1.740 | 1.980 | 0.001854 | 251.80 | 0.93 |
| 80.514 | 1.750 | 1.989 | 0.001833 | 249.60 | 0.93 |
| 81.514 | 1.759 | 1.999 | 0.001842 | 256.30 | 0.90 |
| 82.513 | 1.769 | 2.008 | 0.001845 | 294.40 | 0.69 |
| 84.513 | 1.788 | 2.026 | 0.001848 | 230.60 | 0.92 |
| 85.514 | 1.797 | 2.035 | 0.001861 | 268.10 | 0.80 |
| 86.513 | 1.807 | 2.043 | 0.001863 | 219.30 | 0.86 |
| 87.513 | 1.817 | 2.052 | 0.001861 | 211.40 | 0.86 |
| 89.508 | 1.836 | 2.069 | 0.001874 | 213.20 | 0.96 |
| 90.508 | 1.846 | 2.077 | 0.001874 | 195.20 | 1.06 |
| 91.508 | 1.856 | 2.085 | 0.001902 | 206.50 | 0.94 |
| 92.508 | 1.865 | 2.093 | 0.001917 | 194.70 | 0.94 |
| 93.509 | 1.875 | 2.101 | 0.001962 | 221.00 | 0.87 |
| 94.509 | 1.885 | 2.109 | 0.001980 | 184.70 | 0.96 |
| 95.509 | 1.895 | 2.117 | 0.002002 | 187.90 | 0.89 |
| 96.484 | 1.905 | 2.125 | 0.002007 | 231.40 | 0.79 |
| 97.484 | 1.915 | 2.132 | 0.002002 | 279.70 | 0.69 |
| 98.484 | 1.925 | 2.140 | 0.001985 | 239.10 | 0.73 |
| 99.484 | 1.935 | 2.147 | 0.001959 | 237.70 | 0.75 |
| 100.484 | 1.945 | 2.155 | 0.001918 | 256.60 | 0.77 |
| 101.484 | 1.955 | 2.162 | 0.001874 | 218.20 | 0.87 |
| 102.484 | 1.965 | 2.169 | 0.001846 | 253.30 | 0.84 |
| 103.484 | 1.976 | 2.177 | 0.001808 | 264.80 | 0.83 |
| 104.484 | 1.986 | 2.184 | 0.001782 | 200.90 | 1.07 |
| 105.480 | 1.996 | 2.191 | 0.001762 | 227.50 | 1.02 |
| 106.480 | 2.006 | 2.198 | 0.001763 | 221.20 | 1.02 |
| 107.480 | 2.016 | 2.205 | 0.001770 | 172.30 | 1.23 |

| ΔT | r | Δ | g | Q | δQ |
|---|---|---|---|---|---|
| 108.480 | 2.027 | 2.212 | 0.001770 | 186.70 | 1.11 |
| 109.480 | 2.037 | 2.219 | 0.001808 | 167.50 | 1.30 |
| 110.480 | 2.047 | 2.226 | 0.002007 | 138.50 | 1.22 |
| 111.480 | 2.058 | 2.232 | 0.001826 | 128.20 | 1.46 |
| 112.480 | 2.068 | 2.239 | 0.001840 | 147.10 | 1.42 |
| 113.456 | 2.078 | 2.246 | 0.001868 | 130.60 | 1.41 |
| 114.456 | 2.089 | 2.253 | 0.001876 | 139.30 | 1.26 |
| 115.456 | 2.099 | 2.259 | 0.001908 | 153.50 | 1.31 |
| 116.456 | 2.109 | 2.266 | 0.001937 | 118.30 | 1.43 |
| 117.456 | 2.120 | 2.273 | 0.001957 | 130.50 | 1.35 |
| 118.456 | 2.130 | 2.280 | 0.001969 | 136.20 | 1.22 |
| 119.456 | 2.141 | 2.286 | 0.001989 | 121.60 | 1.29 |
| 120.456 | 2.151 | 2.293 | 0.001990 | 132.00 | 1.20 |
| 121.456 | 2.162 | 2.300 | 0.001988 | 146.80 | 1.15 |
| 122.456 | 2.172 | 2.307 | 0.001964 | 139.40 | 1.13 |

ΔT: Time from perihelion on 30.1 Jan 2015 UT in days
r : Heliocentric distance (AU)
Δ: Geocentric distance (AU)
g: Solar Lyman-α g-factor (photons s$^{-1}$) at 1 AU
Q: Daily-Average Water production rates (molecules s$^{-1}$) from the TRM
δQ: internal 1-sigma uncertainties

Table S7. SOHO/SWAN Observations of C/2015 G2 (MASTER) and Water Production Rates

| ΔT (days) | r (AU) | Δ (AU) | g (s$^{-1}$) | Q (10$^{27}$ molecules s$^{-1}$) | δQ (10$^{27}$ molecules s$^{-1}$) |
|---|---|---|---|---|---|
| -45.896 | 1.159 | 1.464 | 0.002247 | 41.54 | 3.03 |
| -44.896 | 1.146 | 1.430 | 0.001572 | 35.16 | 8.20 |
| -42.895 | 1.121 | 1.363 | 0.002174 | 34.95 | 3.25 |
| -41.895 | 1.108 | 1.329 | 0.002139 | 27.81 | 2.50 |
| -40.895 | 1.096 | 1.295 | 0.002116 | 31.87 | 2.05 |
| -39.895 | 1.083 | 1.261 | 0.002085 | 40.61 | 1.94 |
| -38.895 | 1.071 | 1.227 | 0.002064 | 44.29 | 2.07 |
| -37.843 | 1.058 | 1.191 | 0.002053 | 31.33 | 2.30 |
| -36.843 | 1.046 | 1.156 | 0.002047 | 33.31 | 1.00 |
| -35.843 | 1.034 | 1.122 | 0.001540 | 33.50 | 5.48 |
| -34.837 | 1.022 | 1.087 | 0.001540 | 31.59 | 1.96 |
| -33.838 | 1.011 | 1.052 | 0.002024 | 30.46 | 1.95 |
| -32.837 | 0.999 | 1.018 | 0.002004 | 38.64 | 0.99 |
| -31.818 | 0.988 | 0.982 | 0.001978 | 37.93 | 1.19 |
| -29.809 | 0.965 | 0.914 | 0.001951 | 45.90 | 1.04 |
| -28.809 | 0.955 | 0.880 | 0.001971 | 67.71 | 1.18 |
| -27.790 | 0.944 | 0.845 | 0.002004 | 67.40 | 1.17 |
| -26.790 | 0.934 | 0.812 | 0.002034 | 50.14 | 1.08 |
| -25.780 | 0.923 | 0.779 | 0.001498 | 60.55 | 6.70 |
| -24.780 | 0.913 | 0.747 | 0.002086 | 52.16 | 0.75 |
| -23.760 | 0.904 | 0.715 | 0.002111 | 50.16 | 0.65 |
| -22.751 | 0.894 | 0.684 | 0.002125 | 43.88 | 2.65 |
| -21.751 | 0.885 | 0.654 | 0.002053 | 44.67 | 2.06 |
| -20.731 | 0.876 | 0.625 | 0.001999 | 42.66 | 3.73 |
| -19.723 | 0.868 | 0.598 | 0.001945 | 48.22 | 3.23 |
| -18.703 | 0.859 | 0.573 | 0.001450 | 47.06 | 3.07 |
| -17.694 | 0.851 | 0.549 | 0.001875 | 47.97 | 0.87 |
| -16.673 | 0.844 | 0.528 | 0.001835 | 49.62 | 2.94 |
| -15.666 | 0.837 | 0.510 | 0.001801 | 57.60 | 2.89 |
| -14.644 | 0.830 | 0.495 | 0.001800 | 51.33 | 0.84 |
| -13.638 | 0.823 | 0.483 | 0.001802 | 58.21 | 2.36 |
| -12.615 | 0.817 | 0.475 | 0.001806 | 59.77 | 0.51 |
| -11.610 | 0.812 | 0.470 | 0.001858 | 58.47 | 1.34 |
| -10.611 | 0.807 | 0.470 | 0.001818 | 57.34 | 5.63 |
| -9.584 | 0.802 | 0.474 | 0.001822 | 68.35 | 0.43 |
| -8.166 | 0.796 | 0.485 | 0.001812 | 64.12 | 3.96 |
| -7.139 | 0.792 | 0.498 | 0.001837 | 79.84 | 0.44 |

| ΔT | r | Δ | g | Q | δQ |
|---|---|---|---|---|---|
| -6.140 | 0.789 | 0.514 | 0.001833 | 64.35 | 4.07 |
| -5.136 | 0.786 | 0.533 | 0.001873 | 64.34 | 3.43 |
| -4.112 | 0.784 | 0.555 | 0.001881 | 75.77 | 17.95 |
| -3.112 | 0.782 | 0.579 | 0.001894 | 49.12 | 4.74 |
| -2.107 | 0.781 | 0.605 | 0.001899 | 50.22 | 0.81 |
| 0.922 | 0.780 | 0.693 | 0.001864 | 53.12 | 1.05 |
| 1.945 | 0.781 | 0.725 | 0.001842 | 53.05 | 1.17 |
| 2.945 | 0.782 | 0.758 | 0.001829 | 42.29 | 1.31 |
| 3.951 | 0.784 | 0.791 | 0.001366 | 53.78 | 2.27 |
| 4.952 | 0.786 | 0.824 | 0.001730 | 61.19 | 1.19 |
| 5.974 | 0.788 | 0.859 | 0.001682 | 49.51 | 1.71 |
| 6.974 | 0.791 | 0.894 | 0.001659 | 77.74 | 1.29 |
| 7.981 | 0.795 | 0.928 | 0.001644 | 64.07 | 1.61 |
| 8.981 | 0.799 | 0.963 | 0.001641 | 59.95 | 1.61 |
| 10.002 | 0.804 | 0.999 | 0.001648 | 67.30 | 1.32 |
| 11.002 | 0.809 | 1.033 | 0.001645 | 57.05 | 1.64 |
| 27.092 | 0.937 | 1.566 | 0.001444 | 56.89 | 4.12 |
| 28.092 | 0.947 | 1.596 | 0.001446 | 58.36 | 3.13 |
| 29.087 | 0.958 | 1.626 | 0.001451 | 64.98 | 4.20 |
| 30.597 | 0.974 | 1.670 | 0.001457 | 69.33 | 2.46 |
| 31.764 | 0.987 | 1.704 | 0.001460 | 64.81 | 3.45 |
| 32.931 | 1.000 | 1.737 | 0.001464 | 70.42 | 3.03 |
| 34.098 | 1.014 | 1.769 | 0.001468 | 68.07 | 3.67 |

ΔT: Time from perihelion on 23.8 May 2015 UT in days
r : Heliocentric distance (AU)
Δ: Geocentric distance (AU)
g: Solar Lyman-α g-factor (photons s$^{-1}$) at 1 AU
Q: Daily-Average Water production rates (molecules s$^{-1}$) from the TRM
δQ: internal 1-sigma uncertainties

Table S8. SOHO/SWAN Observations of C/2014 Q1 (PanSTARRS) and Water Production Rates

| ΔT (days) | r (AU) | Δ (AU) | g (s$^{-1}$) | Q (10$^{27}$ molecules s$^{-1}$) | δQ (10$^{27}$ molecules s$^{-1}$) |
|---|---|---|---|---|---|
| -46.608 | 1.189 | 1.988 | 0.0018 | 26.93 | 3.62 |
| -45.608 | 1.170 | 1.966 | 0.0018 | 22.96 | 3.90 |
| -42.608 | 1.112 | 1.900 | 0.0018 | 30.84 | 2.94 |
| -40.589 | 1.072 | 1.856 | 0.0018 | 18.78 | 4.12 |
| -39.590 | 1.052 | 1.835 | 0.0018 | 44.69 | 2.07 |
| -38.589 | 1.032 | 1.813 | 0.0018 | 36.38 | 2.23 |
| -37.589 | 1.012 | 1.792 | 0.0018 | 45.05 | 1.76 |
| -36.589 | 0.992 | 1.771 | 0.0018 | 27.24 | 3.30 |
| -34.589 | 0.952 | 1.729 | 0.0018 | 20.23 | 4.36 |
| -12.216 | 0.477 | 1.350 | 0.0018 | 276.80 | 1.49 |
| -11.050 | 0.453 | 1.337 | 0.0018 | 403.70 | 1.81 |
| -9.883 | 0.430 | 1.325 | 0.0018 | 514.60 | 1.31 |
| -7.550 | 0.388 | 1.303 | 0.0017 | 703.20 | 1.00 |
| -6.106 | 0.365 | 1.290 | 0.0016 | 773.80 | 1.10 |
| -4.939 | 0.349 | 1.280 | 0.0016 | 1854.00 | 0.98 |
| -3.773 | 0.335 | 1.271 | 0.0015 | 2610.00 | 0.59 |
| 13.076 | 0.495 | 1.182 | 0.0017 | 506.40 | 0.40 |
| 14.286 | 0.520 | 1.183 | 0.0017 | 224.50 | 0.52 |
| 16.720 | 0.572 | 1.189 | 0.0017 | 111.90 | 1.03 |
| 17.928 | 0.598 | 1.195 | 0.0017 | 93.75 | 1.00 |
| 19.139 | 0.625 | 1.201 | 0.0017 | 98.41 | 1.10 |
| 20.357 | 0.651 | 1.209 | 0.0017 | 98.86 | 1.12 |
| 21.565 | 0.677 | 1.217 | 0.0017 | 55.50 | 1.65 |
| 22.773 | 0.703 | 1.227 | 0.0017 | 45.13 | 2.21 |
| 26.404 | 0.781 | 1.262 | 0.0017 | 20.72 | 4.23 |
| 27.621 | 0.807 | 1.275 | 0.0017 | 36.32 | 3.02 |
| 28.829 | 0.832 | 1.290 | 0.0017 | 30.53 | 3.17 |
| 30.037 | 0.858 | 1.305 | 0.0017 | 21.56 | 3.97 |
| 31.252 | 0.883 | 1.321 | 0.0017 | 14.22 | 6.30 |
| 33.668 | 0.933 | 1.354 | 0.0017 | 22.12 | 3.77 |
| 34.885 | 0.958 | 1.372 | 0.0017 | 14.72 | 5.93 |

ΔT: Time from perihelion on 6.5 Jul 2015 UT in days
r : Heliocentric distance (AU)
Δ: Geocentric distance (AU)
g: Solar Lyman-α g-factor (photons s$^{-1}$) at 1 AU
Q: Daily-Average Water production rates (molecules s$^{-1}$) from the TRM

δQ: internal 1-sigma uncertainties

Table S9. SOHO/SWAN Observations of C/2013 X1 (PanSTARRS) and Water Production Rates

| ΔT (days) | r (AU) | Δ (AU) | g (s$^{-1}$) | Q (10$^{27}$ molecules s$^{-1}$) | δQ (10$^{27}$ molecules s$^{-1}$) |
|---|---|---|---|---|---|
| -127.769 | 2.235 | 1.615 | 0.001525 | 23.92 | 8.41 |
| -125.769 | 2.214 | 1.634 | 0.001525 | 52.89 | 3.94 |
| -117.781 | 2.131 | 1.723 | 0.001523 | 78.16 | 2.92 |
| -116.781 | 2.120 | 1.736 | 0.001523 | 42.87 | 4.59 |
| -115.781 | 2.110 | 1.748 | 0.001512 | 69.42 | 3.15 |
| -109.796 | 2.049 | 1.829 | 0.001513 | 71.15 | 3.69 |
| -108.796 | 2.039 | 1.843 | 0.001513 | 36.63 | 6.41 |
| -107.795 | 2.029 | 1.857 | 0.001513 | 49.30 | 4.40 |
| -106.795 | 2.018 | 1.871 | 0.001513 | 67.23 | 4.43 |
| -105.795 | 2.008 | 1.885 | 0.001513 | 51.93 | 4.68 |
| -104.795 | 1.998 | 1.899 | 0.001513 | 101.40 | 2.43 |
| -103.795 | 1.988 | 1.913 | 0.001514 | 93.14 | 2.97 |
| -102.795 | 1.978 | 1.928 | 0.001514 | 77.37 | 3.09 |
| -101.795 | 1.968 | 1.942 | 0.001514 | 100.60 | 2.60 |
| -100.795 | 1.958 | 1.956 | 0.001514 | 74.96 | 3.48 |
| -99.811 | 1.948 | 1.970 | 0.001514 | 164.90 | 1.79 |
| -98.794 | 1.938 | 1.984 | 0.001514 | 114.40 | 2.69 |
| -97.811 | 1.928 | 1.998 | 0.001514 | 109.00 | 2.50 |
| -96.812 | 1.919 | 2.012 | 0.001515 | 120.40 | 2.48 |
| -95.812 | 1.909 | 2.026 | 0.001515 | 60.18 | 4.65 |
| -94.812 | 1.899 | 2.040 | 0.001502 | 131.30 | 2.38 |
| -93.812 | 1.889 | 2.053 | 0.001502 | 165.80 | 2.39 |
| -92.812 | 1.879 | 2.067 | 0.001502 | 95.00 | 2.95 |
| -91.812 | 1.870 | 2.080 | 0.001502 | 109.80 | 2.68 |
| -90.812 | 1.860 | 2.094 | 0.001503 | 94.96 | 3.16 |
| -89.812 | 1.850 | 2.107 | 0.001503 | 86.47 | 4.01 |
| -88.812 | 1.841 | 2.120 | 0.001503 | 107.60 | 2.74 |
| -87.822 | 1.831 | 2.132 | 0.001503 | 148.40 | 2.31 |
| -86.822 | 1.822 | 2.145 | 0.001503 | 116.40 | 2.92 |
| -85.822 | 1.812 | 2.157 | 0.001503 | 97.65 | 3.26 |
| -84.822 | 1.803 | 2.170 | 0.001504 | 104.30 | 3.14 |
| -83.821 | 1.793 | 2.182 | 0.001504 | 86.85 | 3.93 |
| -82.821 | 1.784 | 2.194 | 0.001504 | 133.50 | 2.68 |
| -81.821 | 1.775 | 2.205 | 0.001504 | 141.20 | 2.50 |
| -80.821 | 1.765 | 2.216 | 0.001504 | 124.10 | 2.82 |
| -79.094 | 1.749 | 2.235 | 0.001492 | 147.90 | 2.61 |
| -78.093 | 1.740 | 2.246 | 0.001492 | 154.20 | 2.29 |
| -77.093 | 1.731 | 2.257 | 0.001492 | 175.10 | 2.36 |
| -76.093 | 1.722 | 2.267 | 0.001492 | 180.30 | 2.12 |
| -75.093 | 1.713 | 2.277 | 0.001492 | 158.20 | 2.24 |
| -74.093 | 1.704 | 2.286 | 0.001492 | 130.30 | 2.89 |
| -73.093 | 1.695 | 2.295 | 0.001493 | 150.50 | 2.40 |
| -72.115 | 1.687 | 2.304 | 0.001493 | 146.10 | 2.51 |
| -71.115 | 1.678 | 2.313 | 0.001493 | 173.30 | 2.21 |
| -70.115 | 1.669 | 2.321 | 0.001493 | 130.30 | 2.88 |
| -69.115 | 1.660 | 2.330 | 0.001482 | 171.10 | 2.54 |
| -68.115 | 1.652 | 2.337 | 0.001482 | 108.10 | 3.94 |
| -66.092 | 1.635 | 2.352 | 0.001482 | 156.30 | 3.18 |
| -65.116 | 1.626 | 2.359 | 0.001482 | 139.00 | 2.81 |
| -64.116 | 1.618 | 2.365 | 0.001483 | 137.80 | 3.27 |
| -63.116 | 1.610 | 2.371 | 0.001483 | 196.50 | 2.14 |
| -62.116 | 1.602 | 2.377 | 0.001483 | 146.60 | 3.28 |
| -61.116 | 1.594 | 2.382 | 0.001472 | 156.70 | 3.08 |
| -60.116 | 1.586 | 2.387 | 0.001472 | 129.30 | 3.50 |
| -59.116 | 1.578 | 2.392 | 0.001472 | 144.40 | 2.83 |
| -58.116 | 1.570 | 2.396 | 0.001473 | 155.50 | 3.53 |
| -57.116 | 1.562 | 2.401 | 0.001473 | 248.00 | 2.64 |
| -56.116 | 1.554 | 2.404 | 0.001473 | 196.50 | 4.04 |
| -43.788 | 1.467 | 2.417 | 0.001456 | 122.80 | 5.98 |
| -16.734 | 1.338 | 2.222 | 0.001413 | 326.70 | 1.94 |
| -15.734 | 1.335 | 2.209 | 0.001414 | 367.60 | 1.52 |
| -14.733 | 1.333 | 2.196 | 0.001414 | 441.10 | 1.10 |
| -13.733 | 1.330 | 2.182 | 0.001414 | 356.80 | 1.40 |
| -12.733 | 1.328 | 2.167 | 0.001410 | 311.90 | 1.53 |
| -11.733 | 1.326 | 2.152 | 0.001410 | 324.00 | 1.44 |
| -10.733 | 1.324 | 2.137 | 0.001411 | 283.10 | 1.35 |
| -9.735 | 1.322 | 2.121 | 0.001407 | 191.70 | 1.92 |
| -7.735 | 1.319 | 2.089 | 0.001407 | 139.50 | 2.76 |
| -6.735 | 1.318 | 2.072 | 0.001404 | 104.10 | 2.95 |
| -5.735 | 1.317 | 2.055 | 0.001404 | 144.60 | 2.15 |
| -4.735 | 1.316 | 2.037 | 0.001404 | 114.80 | 3.07 |
| -3.735 | 1.315 | 2.019 | 0.001404 | 148.90 | 7.10 |
| -1.735 | 1.315 | 1.982 | 0.001403 | 106.80 | 3.20 |
| -0.736 | 1.314 | 1.962 | 0.001403 | 150.10 | 2.34 |
| 0.264 | 1.314 | 1.943 | 0.001403 | 111.20 | 2.69 |
| 1.264 | 1.314 | 1.923 | 0.001404 | 144.90 | 2.07 |
| 2.264 | 1.315 | 1.903 | 0.001404 | 176.10 | 2.04 |
| 3.264 | 1.315 | 1.882 | 0.001404 | 99.80 | 2.82 |
| 4.264 | 1.316 | 1.861 | 0.001402 | 102.50 | 2.94 |
| 5.264 | 1.317 | 1.840 | 0.001402 | 135.70 | 1.99 |
| 6.264 | 1.318 | 1.818 | 0.001403 | 113.10 | 2.23 |
| 7.249 | 1.319 | 1.797 | 0.001402 | 108.50 | 2.43 |
| 8.249 | 1.320 | 1.775 | 0.001402 | 152.50 | 1.95 |
| 9.249 | 1.322 | 1.752 | 0.001402 | 106.20 | 2.27 |
| 11.249 | 1.325 | 1.706 | 0.001402 | 131.70 | 1.81 |
| 13.249 | 1.329 | 1.660 | 0.001403 | 132.20 | 2.05 |
| 14.250 | 1.332 | 1.635 | 0.001402 | 140.60 | 1.96 |
| 15.544 | 1.335 | 1.604 | 0.001403 | 117.70 | 1.45 |
| 17.250 | 1.339 | 1.562 | 0.001403 | 120.30 | 1.53 |
| 18.250 | 1.342 | 1.537 | 0.001404 | 142.70 | 2.08 |
| 19.250 | 1.346 | 1.512 | 0.001404 | 134.50 | 1.67 |
| 20.250 | 1.349 | 1.487 | 0.001405 | 146.20 | 1.70 |
| 21.250 | 1.352 | 1.461 | 0.001406 | 146.50 | 1.87 |
| 22.250 | 1.356 | 1.436 | 0.001406 | 141.10 | 1.89 |
| 24.251 | 1.363 | 1.384 | 0.001406 | 168.60 | 1.79 |
| 30.252 | 1.390 | 1.227 | 0.001413 | 131.50 | 1.75 |
| 31.252 | 1.395 | 1.201 | 0.001413 | 127.20 | 1.53 |
| 35.252 | 1.416 | 1.096 | 0.001417 | 142.30 | 0.76 |
| 40.299 | 1.445 | 0.968 | 0.001424 | 118.60 | 14.48 |
| 41.299 | 1.451 | 0.943 | 0.001424 | 109.70 | 7.33 |
| 42.311 | 1.457 | 0.919 | 0.001425 | 105.20 | 9.57 |
| 46.092 | 1.482 | 0.833 | 0.001432 | 72.46 | 2.88 |
| 47.083 | 1.489 | 0.812 | 0.001432 | 54.27 | 3.41 |
| 48.083 | 1.495 | 0.791 | 0.001432 | 64.64 | 3.09 |
| 49.065 | 1.502 | 0.772 | 0.001439 | 22.07 | 7.85 |
| 50.065 | 1.509 | 0.753 | 0.001439 | 67.37 | 3.93 |
| 51.064 | 1.517 | 0.736 | 0.001440 | 65.49 | 2.94 |
| 52.058 | 1.524 | 0.720 | 0.001440 | 44.55 | 3.54 |
| 53.058 | 1.531 | 0.705 | 0.001440 | 40.10 | 3.95 |
| 54.059 | 1.539 | 0.691 | 0.001441 | 53.03 | 4.13 |
| 55.037 | 1.546 | 0.679 | 0.001449 | 64.95 | 3.33 |
| 56.037 | 1.554 | 0.669 | 0.001450 | 47.17 | 3.35 |
| 59.031 | 1.577 | 0.648 | 0.001451 | 49.53 | 2.64 |
| 60.032 | 1.585 | 0.645 | 0.001451 | 38.40 | 4.03 |
| 61.032 | 1.593 | 0.643 | 0.001451 | 28.92 | 6.40 |
| 62.006 | 1.601 | 0.644 | 0.001460 | 31.03 | 4.27 |
| 63.006 | 1.609 | 0.647 | 0.001461 | 54.30 | 2.55 |
| 77.247 | 1.733 | 0.866 | 0.001481 | 30.86 | 6.35 |
| 78.248 | 1.742 | 0.891 | 0.001482 | 60.45 | 3.86 |
| 101.209 | 1.962 | 1.572 | 0.001504 | 35.35 | 7.63 |
| 102.222 | 1.972 | 1.605 | 0.001505 | 27.72 | 12.55 |

ΔT: Time from perihelion on 20.8 Apr 2016 UT in days r : Heliocentric distance (AU)  
Δ: Geocentric distance (AU)  
g: Solar Lyman-α g-factor (photons s$^{-1}$) at 1 AU  
Q: Daily-Average Water production rates (molecules s$^{-1}$) from the TRM  
δQ: internal 1-sigma uncertainties